\documentclass[aps,prl,twocolumn,superscriptaddress,english]{revtex4}
\usepackage{amssymb}
\usepackage{graphicx}
\usepackage{amsmath}
\usepackage{amsthm}
\usepackage{amsfonts}
\usepackage[T1]{fontenc}
\usepackage[latin9]{inputenc}
\usepackage{array}
\usepackage{multirow}
\usepackage{color}
\usepackage{esint}
\usepackage{bm}
\usepackage{color}
\usepackage{bbm}
\usepackage{hyperref}
\usepackage{babel}
\usepackage{titlesec}

\begin{document}

\global\long\def\id{\mathbbm{1}}
\global\long\def\ui{\mathbbm{i}}
\global\long\def\ud{\mathrm{d}}

\title{Dynamical detection of topological charges}

\author{Long Zhang}
\affiliation{International Center for Quantum Materials, School of Physics, Peking University, Beijing 100871, China.}
\affiliation{Collaborative Innovation Center of Quantum Matter, Beijing 100871, China.}

\author{Lin Zhang}
\affiliation{International Center for Quantum Materials, School of Physics, Peking University, Beijing 100871, China.}
\affiliation{Collaborative Innovation Center of Quantum Matter, Beijing 100871, China.}

\author{Xiong-Jun Liu}
\thanks{Corresponding author: xiongjunliu@pku.edu.cn}
\affiliation{International Center for Quantum Materials, School of Physics, Peking University, Beijing 100871, China.}
\affiliation{Collaborative Innovation Center of Quantum Matter, Beijing 100871, China.}

\begin{abstract}
We propose a generic scheme to characterize topological phases via detecting topological charges by quench dynamics.
A topological charge is defined as the chirality of a monopole at Dirac or Weyl point of spin-orbit field, and topological phases can
be classified by total charges in the region enclosed by the so-called band-inversion surfaces (BISs).
We show that both the topological monopoles and BISs can be identified by non-equilibrium spin dynamics caused by a sequence of quenches.
From an emergent dynamical field given by time-averaged spin textures, the topological charges, as well as the topological invariant, can be readily obtained. An explicit advantage in this scheme is that only a single spin component needs to be measured to detect all the information of the topological phase.
We numerically examine two realistic models, and propose a feasible experimental setup for the measurement. This work opens a new
way to dynamically classify topological phases.
\end{abstract}
\maketitle

{\em Introduction.---}
Topological quantum matter~\cite{Hasan2010,Qi2011} has ignited extensive research in recent years.
In particular, experimental advances in cold atoms have promoted the realization of
novel topological states including the one-dimensional (1D) Su-Schrieffer-Heeger (SSH) chain~\cite{Su1980,Atala2013},
1D chiral topological phase~\cite{Liu2013,Song2018} and 2D Chern insulators~\cite{Aidelsburger2013,Miyake2013,Jotzu2014,Aidelsburger2015,Wu2016,Sun2017}.
Compared with solid state materials,
ultracold atoms are ideal platforms with high controllability, which makes it accessible to
experimentally investigate non-equilibrium dynamics caused by quenches~\cite{Song2018,quench1,quench2,quench3,quench4,quench5,Flaschner2016,Flaschner2018}.

Very recently, several works~\cite{Wang2017,Zhang2018,Sun2018,Tarnowski2017,Yu2018} have focused on characterizing
topology of Hamiltonian by quench dynamics. In particular, a non-equilibrium classification of topological states,
which in equilibrium are characterized by integer invariants,
is established and shows experimental feasibility to detect bulk topology with high precision~\cite{Zhang2018}.
It is shown that the bulk topology of the post-quench phase can be classified by
a dynamical topological invariant defined on the so-called band-inversion surfaces (BISs)~\cite{Zhang2018}.
This classification theory has been applied in a latest experiment based on 2D Chern insulator~\cite{Sun2018},
which shows that the dynamical measurement of topological states has a much higher precision over the equilibrium measurement strategies~\cite{Sun2017}.
Applications to dynamical topological phase transition~\cite{WYi2018} and non-Hermitian topological phases~\cite{Zhou2018,Qiu2018,KWang2018} are also considered.

In this letter, we propose a new scheme of dynamical classification to characterize topology by detecting topological charges.
The topological charge is defined as the chirality of a monopole (vanishing point) of the spin-orbit (SO) field, and the topology can
be determined by the total charges in the region enclosed by BISs. The key idea is that through a sequence of quenches
with respect to all (pseudo)spin quantization axes, both the BISs and topological charges of monopoles can be directly identified by
measuring the time-averaged spin polarization. 
Unlike our previous dynamical scheme~\cite{Zhang2018} which quenches along a certain (pseudo)spin axis but needs to measure all the (pseudo)spin components, the present new detection method necessitates measuring only a single spin component.
On the other hand, compared with the method of measuring the linking number of trajectories in the momentum-time space~\cite{Wang2017,Tarnowski2017}, which is valid for 2D phases, the present scheme can be applied to characterize topological states of all dimensions.
In the last part of this work, we propose an experimental setup to simulate the dynamical detection on 2D quantum anomalous Hall model, which can be well achieved in experiment.

\begin{figure}
\includegraphics[width=0.48\textwidth]{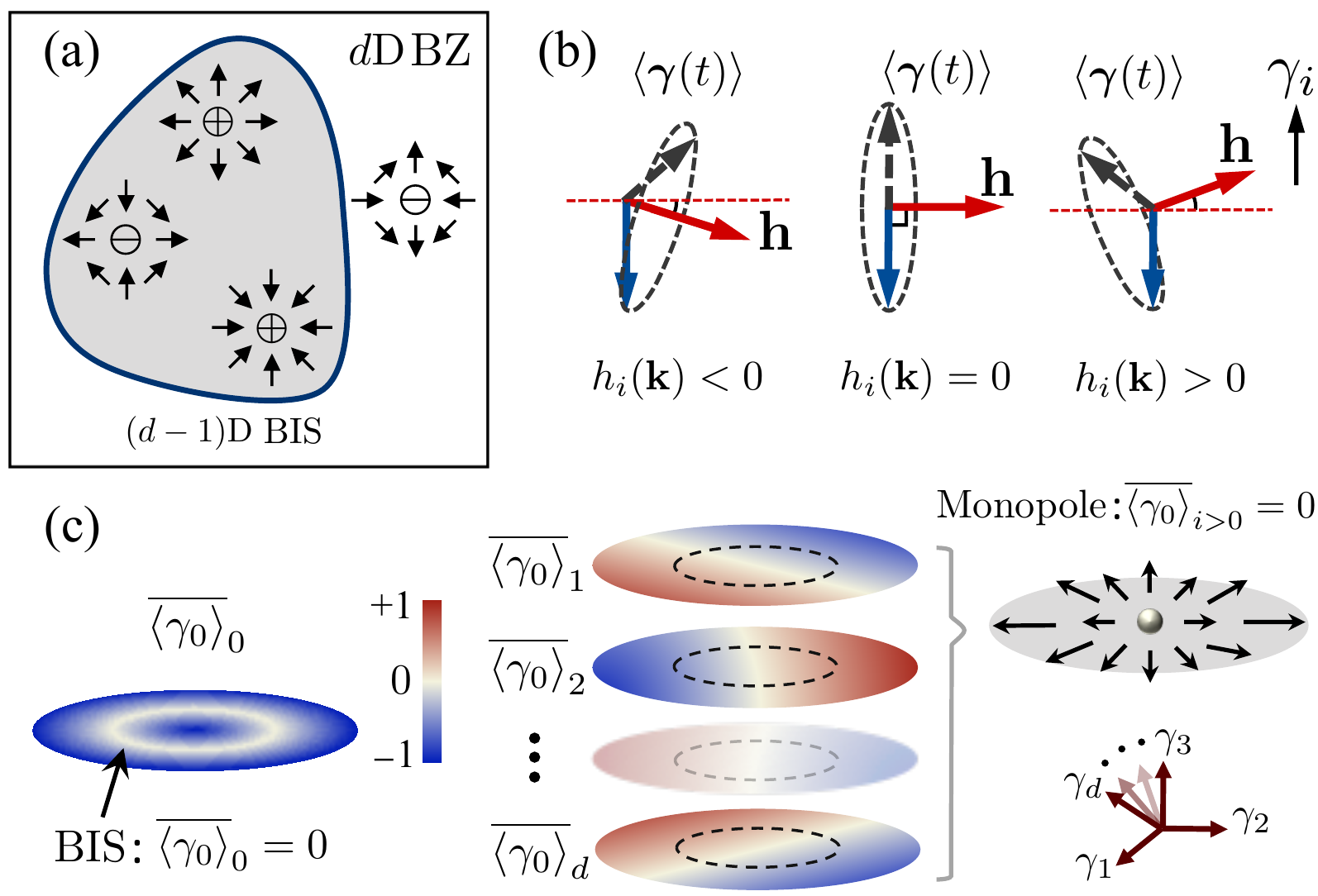}
\caption{Scheme of dynamical detection. (a) Topological phases can
be characterized by the total charges in the region enclosed by BISs.
(b) Quenching $h_{i}$  corresponds to initializing a polarized state (blue arrow) for the spin precession about the post-quench vector field ${\bf h}$ (red arrow).
Time-averaged spin polarization $\overline{\langle{\bm\gamma}\rangle}$ reflect the $h_i$ component:
The surface with $h_i({\bf k})=0$ would exhibit no polarization, while for $h_i({\bf k})<0$ (or $h_i({\bf k})>0$),
the time-averaged spin orientation points to the (or opposite) direction of ${\bf h}$.
(c) A BIS can be identified as the surface with vanishing spin polarization in the time-averaged spin texture $\overline{\langle\gamma_0({\bf k})\rangle}_0$
obtained by quenching $h_0$, which also emerges in $\overline{\langle\gamma_0({\bf k})\rangle}_{i}$ by quenching $h_{i>0}$ (dashed lines).
In spin textures $\overline{\langle\gamma_0({\bf k})\rangle}_{i>0}$, other surfaces besides the BIS exhibiting no polarization are the surfaces $h_{i}({\bf k})=0$.
The topological monopoles are located at the intersections of all the $h_{i>0}({\bf k})=0$ surfaces, and
the charges can be also characterized via the spin textures $\overline{\langle\gamma_0({\bf k})\rangle}_{i}$.
}\label{fig1}
\end{figure}

{\em Generic model.---}
We start with the model of a generic $d$-dimensional (dD) gapped topological phase, which can be insulator or superconductor, and is described by the basic Hamiltonian
\begin{equation}
{\cal H}({\bf k})={\bf h}({\bf k})\cdot{\boldsymbol \gamma}=h_0({\bf k})\gamma_0+\sum_{i=1}^d h_i({\bf k})\gamma_i.
\end{equation}
Here the $\gamma$ matrices define a (pseudo)spin and obey the Clifford algebra $\left\{\gamma_{i},\gamma_{j}\right\} =2\delta_{ij}\mathbbm{1}$
for $i,j=0,1\dots,d$, and the $(d+1)$D vector field ${\bf h}(\mathbf{k})$ describes an effective Zeeman field depending on the momentum $\mathbf{k}$ in the Brillouin zone (BZ)~\cite{Supp}.
The Hilbert space at each ${\bf k}$ are of dimensionality $n_{d}=2^{d/2}$ (or $2^{(d+1)/2}$) if $d$ is even (or odd), involving
the minimal bands to open a topological gap~\cite{Zhang2018,Chiu2013}.
In the 1D and 2D regimes, the $\gamma$ matrices simply reduce to the Pauli matrices, and $\mathcal{H}(\mathbf{k})$ describes a two-band model for the topological states, e.g., the well-known SSH model~\cite{Su1980, Chiu2016} for 1D and the quantum anomalous Hall model~\cite{Haldane1988,XJLiu2014} for 2D. Similarly, for 3D and 4D phases, the $\gamma$ matrices take the Dirac forms, and a fully gapped topological phase has at least four bands~\cite{Schnyder2008,Zhang2001}.
Similar to the convention used in Ref.~\cite{Zhang2018}, we choose $h_{0}(\mathbf{k})$ to characterize the `dispersion' of the $n_{d}$ decoupled bands, and define the remaining components as the SO field ${\bf h}_{\mathrm{so}}(\mathbf{k})\equiv(h_{1},\dots,h_{d})$, which depict the coupling between different bands.

In the previous work~\cite{Zhang2018}, a bulk-surface duality was shown that the $d$D bulk topology
can be characterized by a $(d-1)$D invariant defined on
BISs. A BIS refers to
the $(d-1)$D band-crossing surface with $h_{0}(\mathbf{k})=0$.
The SO field ${\bf h}_{\mathrm{so}}({\bf k})$ opens the gap, and brings out nontrivial topology.
The defined $(d-1)$D topological invariant counts the winding number of the SO field on the BISs.
Analogous to the magnetic field, one can also consider the monopoles of SO field, and
the topological invariant is thus viewed as the flux of the monopoles through the BIS [see Fig.~\ref{fig1}(a)].
It is easily seen that the monopoles are located at ${\bf k}={\bm \varrho}$ where ${\bf h}_{\mathrm{so}}({\bm \varrho})=0$, such that
the gap is closed and then reopened as a monopole passes through a BIS, indicating a topological phase transition.
As detailed in the Supplementary Information~\cite{Supp}, we show that the topological invariant reads
\begin{align}\label{topo_invar}
{\cal W}&=\sum_{i\in{\cal V}_{\rm BIS}}{\cal C}_i,
\end{align}
with
\begin{align}
{\cal C}_i&={\rm sgn}[J_{{\bf h}_{\rm so}}({\bm \varrho}_i)]
\end{align}
being the topological charge of the $i$th monopole
in the region ${\cal V}_{\rm BIS}$ enclosed by BISs. Here $J_{{\bf h}_{\rm so}}({\bf k})\equiv \det\left[(\partial h_{\rm so,i}/\partial k_j)\right]$ is the Jacobian determinant
and ${\cal V}_{\rm BIS}$ denotes the region $h_0({\bf k})<0$.

The invariant in Eq.~(\ref{topo_invar}), as a summation of topological charges, is directly related to the Brouwer degree~\cite{Felsager_book,Milnor_book,Sticlet2012} of
the mapping $\hat{\bf h}({\bf k})\equiv{\bf h}({\bf k})/|{\bf h}({\bf k})|$ from the BZ torus $T^d$ to $d$D spherical surface $S^d$. This formula can be intuitively interpreted
as the effective number of times that the parametric surface ${\cal D}$ traced by the vector ${\bf h}(\bold k)$ passes through the negative $\gamma_0$ axis~\cite{Supp}.
The intersection points are locations ($\bold h_{\rm so}=0$) of monopoles, with the charges $\pm1$ indicating the orientation (or chirality) of the manifold at these intersections. By summing up all the orientation numbers,
we obtain the winding number or Chern number of the $d$D topological phase.

{\em Dynamical detection of the topological charges.---}
We shall show that BISs and topological charges of monopoles can be identified via
quantum spin dynamics induced by, respectively, quenching $h_0$ and a sequence of quenches of $h_{i>0}$.
The basic idea is as follows.
We focus on the time evolution of spin polarization triggered by quenches.
For quenching $h_i$ ($i=0,1,2,\cdots,d$), we initialize a polarized state along this axis,
which is achieved by tuning a large constant magnetization for this component $h_{i}(\mathbf{k})\approx m_i$.
After the quench, the momentum-linked spin $\langle{\bm\gamma}(t)\rangle$
processes around the vector field ${\bf h}(\mathbf{k})$.
In the unitary evolution, the time-averaged spin polarization $\overline{\langle{\bm\gamma}(t)\rangle}$ directly reflects the $h_i$ component [see Fig.~\ref{fig1}(b)].
On the surfaces with $h_i({\bf k})=0$, the spin orientation is always perpendicular to the procession axis,
leading to vanishing polarization. In the region with $h_i({\bf k})<0$ (or $h_i({\bf k})>0$), the vector $\overline{\langle\bm\gamma(t)\rangle}$ is in the (or opposite)
direction of the field ${\bf h}$. Taking $\overline{\langle\gamma_0\rangle}$ as the measurement,
the observations fall into two categories---the first corresponds to quenching $h_0$
and the other corresponds to quenching ${\bf h}_{\rm so}$.
For quenching $h_0$,  one can identify the BISs as the momentum points with vanishing polarization $\overline{\langle\gamma_0\rangle}_0=0$ [see Fig.~\ref{fig1}(c)].
On the other hand, for the second category of quenching $h_{i>0}$, the time-averaged spin polarization $\overline{\langle\gamma_0\rangle}_{i}$ also vanishes on surfaces with $h_{i}({\bf k})=0$.
The topological monopoles, located at the intersections of all the $h_{i>0}({\bf k})=0$ surfaces, are then dynamically identified by the points with $\overline{\langle\gamma_0\rangle}_{i>0}=0$ for all $h_{i>0}$ quenches. The topological charges can further be characterized from a dynamical field constructed by time-averaged spin textures [Fig.~\ref{fig1}(c)], as detailed below.

To be more specific, the system described by density matrix $\rho_i(0)$ is initialized in the ground state of the pre-quench Hamiltonian, with the spins being
fully polarized in the opposite direction of the $\gamma_{i}$ axis.
The quantum dynamics after the quench is governed by the unitary evolution operator $U(t)=\exp(-\ui\mathcal{H}t)$.
The time-dependent density matrix $\rho_i(t)=U(t)\rho_i(0)U^\dagger(t)$ and the spin polarization is given by $\left\langle \gamma_{0}\right\rangle_i={\rm Tr}\left[\rho_i(t)\gamma_0\right]$. After some calculations~\cite{Supp}, the time-averaged spin texture reads
\begin{eqnarray}\label{gamma0i}
\overline{\left\langle \gamma_{0}(\bold k)\right\rangle}_i=-h_{0}(\bold k)h_{i}(\bold k)/E^{2}(\bold k),
\end{eqnarray}
where $E(\bold k)=\bigr[\sum_{i=0}^{d}h_{i}^{2}\bigr]^{1/2}$.
It is interesting that no matter which axis is quenched, the spin texture $\overline{\left\langle \gamma_{0}(\bold k)\right\rangle }_i$ always vanishes on the BISs (with $h_0=0$).
Hence, the BISs are the surfaces with vanishing time-averaged spin polarization independent of the quench way, i.e.
$\mathrm{BISs}=\{\mathbf{k}\vert\forall i,\overline{\left\langle{\gamma_0}(\mathbf{k})\right\rangle}_i=0\}$.
Besides, the Eq.~(\ref{gamma0i}) also gives $\overline{\left\langle \gamma_{0}(\bold k)\right\rangle }_{i}=0$ on the surface
with $h_{i}({\bf k})=0$. Accordingly, the location $\bold k={\bm \varrho}$ of a monopole can be found by $\overline{\left\langle \gamma_{0}({\bm \varrho})\right\rangle }_{i}=0$ for all $i\neq0$ but $\overline{\left\langle \gamma_{0}({\bm \varrho})\right\rangle }_{0}\neq0$. Finally, to characterize the charge,
one notices that near the point ${\bf k}={\bm \varrho}$,
the time-averaged spin texture in quenching $h_i$ can measure the $i$-th component of the SO field as
$\overline{\left\langle \gamma_{0}({\bf k})\right\rangle}_i\big\vert_{{\bf k}\to{\bm \varrho}}=-h_{i}(\bold k)/h_{0}({\bm \varrho}).$
Enlightened by this result, we define a dynamical field ${\bm \Theta}({\bf k})$ to characterize the topological charge, with the components being given by
\begin{equation}\label{fi}
\Theta_i({\bf k})\equiv-\frac{{\rm sgn}[h_0({\bf k})]}{{\cal N}_{\bf k}}\overline{\left\langle \gamma_{0}({\bf k})\right\rangle}_i,
\end{equation}
where ${\cal N}_{\bf k}$ is the normalization factor. It can be shown directly that near the location ${\bm \varrho}$ of the topological monopole, the dynamical field
$\Theta_i({\bf k})\big\vert_{{\bf k}\to{\bm \varrho}}={h}_{{\rm so},i}({\bf k})$.
With this result, we reach finally that the topological charge is dynamically determined by
\begin{equation}
{\cal C}_i={\rm sgn}[J_{{\bm \Theta}}({\bm \varrho}_i)].
\end{equation}

\begin{figure}
\includegraphics[width=0.5\textwidth]{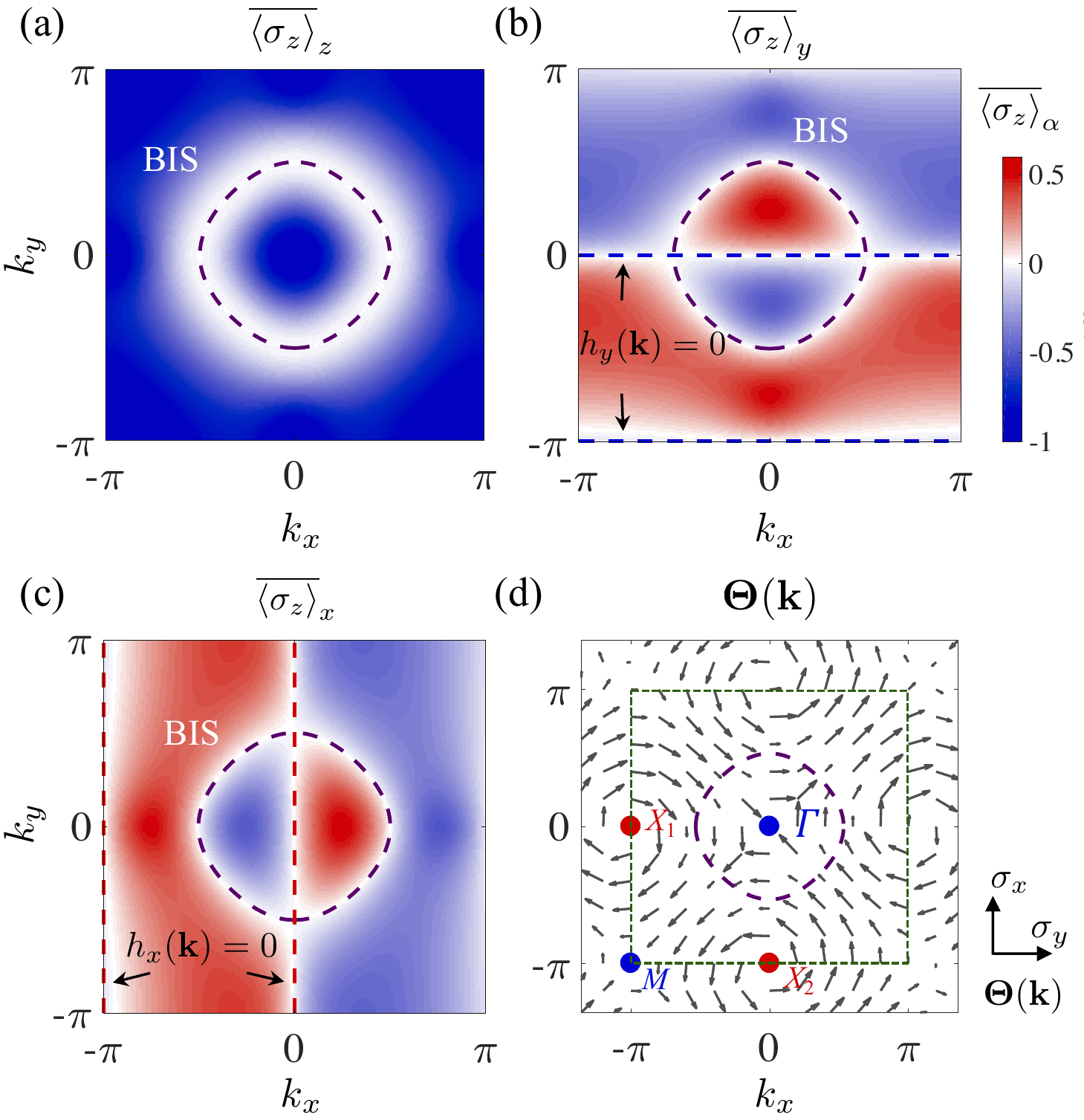}
\caption{Numerical results of 2D QAH model. (a-c) Time-averaged spin textures $\overline{\langle\sigma_z({\bf k})\rangle}$ via quenching $h_z$ (a), $h_y$ (b) and $h_x$ (c).
In (a), the magnetization $m_z$ is quenched from $25t_0$ to $t_0$, with $m_x=m_y=0$; for (b) [or (c)], the quench is varying $m_{y}$ (or $m_{x}$) from $25t_0$ to 0 and $m_z$ from 0 to $t_0$, while keeping $m_x=0$ (or $m_y=0$). Here we take $t_{\rm so}=t_0$.
In addition to the ring-shape structure, which characterizes the BIS, two lines with vanishing polarization emerge in the spin textures $\overline{\langle\sigma_z({\bf k})\rangle}_y$ (b) and $\overline{\langle\sigma_z({\bf k})\rangle}_x$ (c), giving the interfaces with $h_y({\bf k})=0$ and $h_x({\bf k})=0$, respectively.
(d) The dynamical field ${\bm \Theta}({\bf k})$, constructed from the spin textures, characterizes the topological charges at the four monopoles: (blue) ${\cal C}=-1$ at $\it\Gamma$ $(0,0)$ and
$M$ $(-\pi,-\pi)$; (red) ${\cal C}=+1$ at $X_{1}$ $(0,-\pi)$ and $X_{2}$ $(-\pi,0)$. The dashed green line denotes the first BZ.
}\label{fig2}
\end{figure}

{\em Application to two models.---}
We illustrate our scheme by numerally examining two different models.
First, we consider the 2D quantum anomalous Hall (QAH) model ${\cal H}({\bf k})={\bf h}({\bf k})\cdot{\bm \sigma}$~\cite{XJLiu2014,Zhang_book}, where the vector field reads
${\bf h}({\bf k})=(m_x+t_{\rm so}\sin k_x,m_y+t_{\rm so}\sin k_y,m_z-t_0\cos k_x-t_0\cos k_y)$. This model has been realized in cold atom experiments~\cite{Wu2016,Sun2017,Sun2018}.
The bulk topology is determined by $m_z$ ($m_x=m_y=0$) that the Chern number ${\rm Ch}_1=-{\rm sgn}(m_z)$ for $0<|m_z|<2t_0$ and ${\rm Ch}_1=0$ for $|m_{z}|\geq2t_{0}$. We take $h_0\equiv h_z, \bold h_{\rm so}\equiv(h_y,h_x)$~\cite{Supp}, and the quench is performed by suddenly varying $m_\alpha$ ($\alpha=x,y,z$). Note that only the time evolution of spin polarization of the $\sigma_z$-component needs to be measured after each quench process to obtain all the information of topology [see Fig.~\ref{fig2}(a-c)]. The spin textures $\overline{\langle\sigma_z({\bf k})\rangle}_\alpha$ in all three quenches ($\alpha=x,y,z$) show clearly a ring-shape structure, which characterizes the BIS. Besides, spin textures in (b) and (c), respectively, exhibits two lines with vanishing polarization, which indicate the surfaces with $h_y({\bf k})=0$ [for (b)] and $h_x({\bf k})=0$ [for (c)]. These four lines have four intersection points marking the monopoles at $\Gamma, M$ and $X_{1,2}$ points, and the dynamical field ${\bm \Theta}({\bf k})$ obtained by spin textures in (b-c) determines the charge $\pm1$ at each point via
${\rm sgn}[J_{{\bm \Theta}}({\bf k})]$ [see Fig.~~\ref{fig2}(d)].
For $m_z=t_0$, the ring encloses the monopole with ${\cal C}=-1$ at $\Gamma$ point, giving the Chern number ${\rm Ch}_1=-1$.

\begin{figure}
\includegraphics[width=0.5\textwidth]{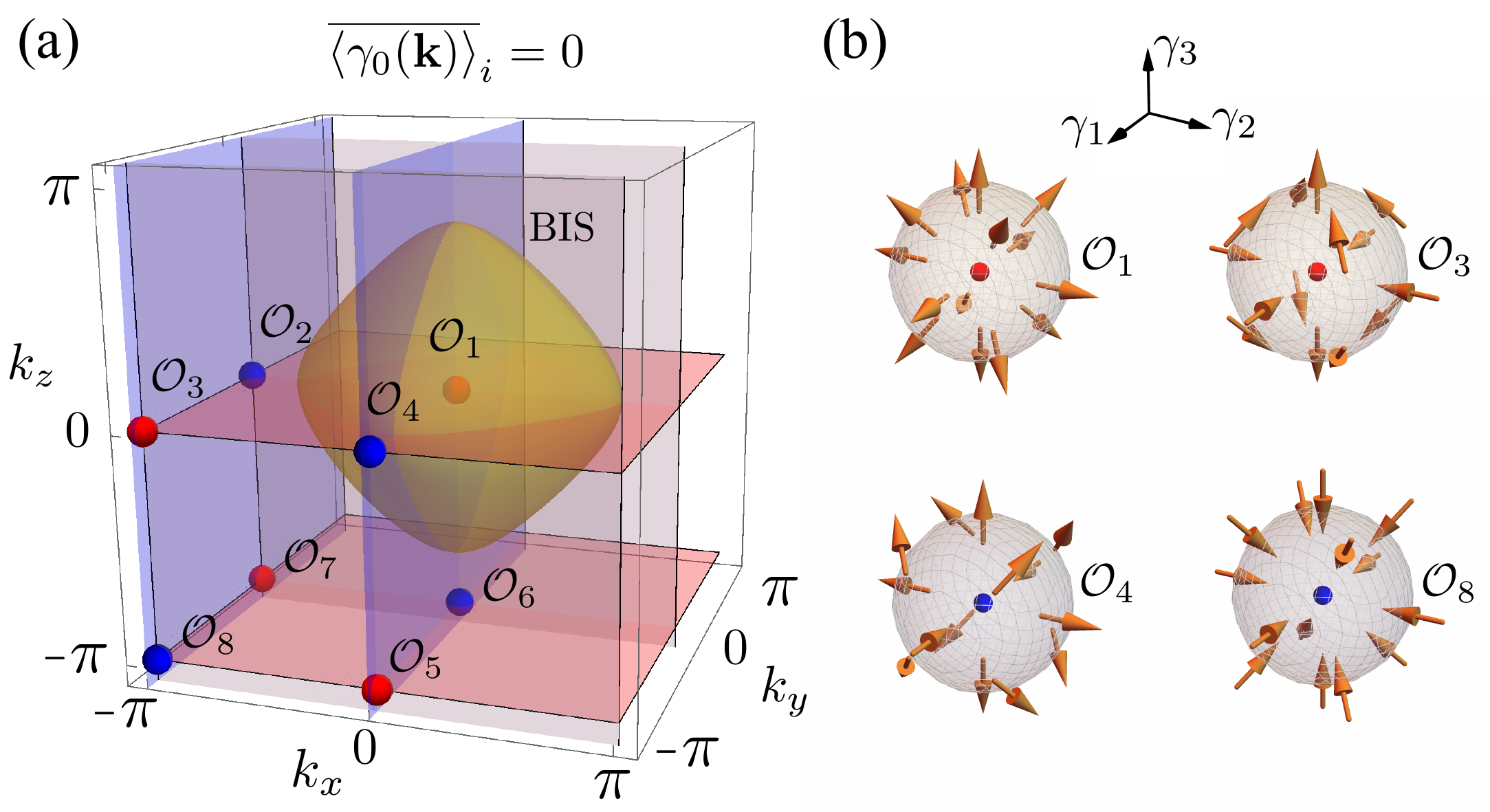}
\caption{Dynamical detection of 3D topological phases.
(a) Time-averaged spin textures determine the BIS with $\overline{\langle\gamma_0({\bf k})\rangle}_0=0$ (green surface)
and identify the monopoles as the intersection points of the surfaces $\overline{\langle\gamma_0({\bf k})\rangle}_{i>0}=0$.
(b) Topological charges can be characterized by the constructed dynamical field, with ${\cal C}=1$ (red) at ${\cal O}_{1,3,5,7}$ and
${\cal C}=-1$ (blue) at ${\cal O}_{2,4,6,8}$.
The details can be found in Ref.~\cite{Supp}.
}\label{fig3}
\end{figure}

We further consider the application to a 3D topological phase, whose Hamiltonian reads $\mathcal{H}(\mathbf{k})={\bf h}(\mathbf{k})\cdot{\bm\gamma}$,
with $h_{0}(\mathbf{k})=m_{0}-t_{0}\sum_{i}\cos k_{r_i}$ and
$h_{i}=m_i+t_{\mathrm{so}}\sin k_{r_i}$ [$(r_1,r_2,r_3)=(x,y,z)$ for $i=1,2,3$]. Here we take
$\gamma_{0}=\sigma_{z}\otimes\tau_{x}$, $\gamma_{1}=\sigma_{x}\otimes{\id}$,
$\gamma_{2}=\sigma_{y}\otimes{\id}$ and $\gamma_{3}=\sigma_{z}\otimes\tau_{z}$, where
$\sigma_{x,y,z}$ and $\tau_{x,y,z}$ are both Pauli matrices.
The topological phases are classified by 3D winding numbers.
The trivial phase corresponds
to $\left|m_{0}\right|>3t_{0}$, while the topological phases include three regions ($m_{i>0}=0$):
(I) $t_{0}<m_{0}<3t_{0}$ with winding number $\nu_3=1$; (II) $-t_{0}<m_{0}<t_{0}$
with $\nu_3=-2$; and (III) $-3t_{0}<m_{0}<-t_{0}$ with $\nu_3=1$.
We observe the time-averaged spin textures $\overline{\langle\gamma_0({\bf k})\rangle}_i$ after quenching different components~\cite{Supp}.
In all cases, the post-quench state takes the parameters $m_0=1.3$, $m_{i>0}=0$ and $t_{\rm so}=t_0$. The pre-quench state is with $m_i=25t_0$, corresponding to quenching $m_i$ ($i=0,1,2,3$). The surface with $\overline{\langle\gamma_0({\bf k})\rangle}_0=0$ characterizes the BIS, and $\overline{\langle\gamma_0({\bf k})\rangle}_{i>0}=0$ locate the topological monopoles as their intersections [see Fig.~\ref{fig3}(a)].
The dynamical field ${\bm \Theta}({\bf k})$, constructed from $\overline{\langle\gamma_0({\bf k})\rangle}_{i}$ by following Eq.~(\ref{fi}),
reflects the charges [Fig.~\ref{fig3}(b)]. One can see that the BIS only surrounds a single monopole with ${\cal C}=+1$,
which reveals that the post-quench state lies in the topological phase with $\nu_3=1$.

\begin{figure}
\includegraphics[width=0.5\textwidth]{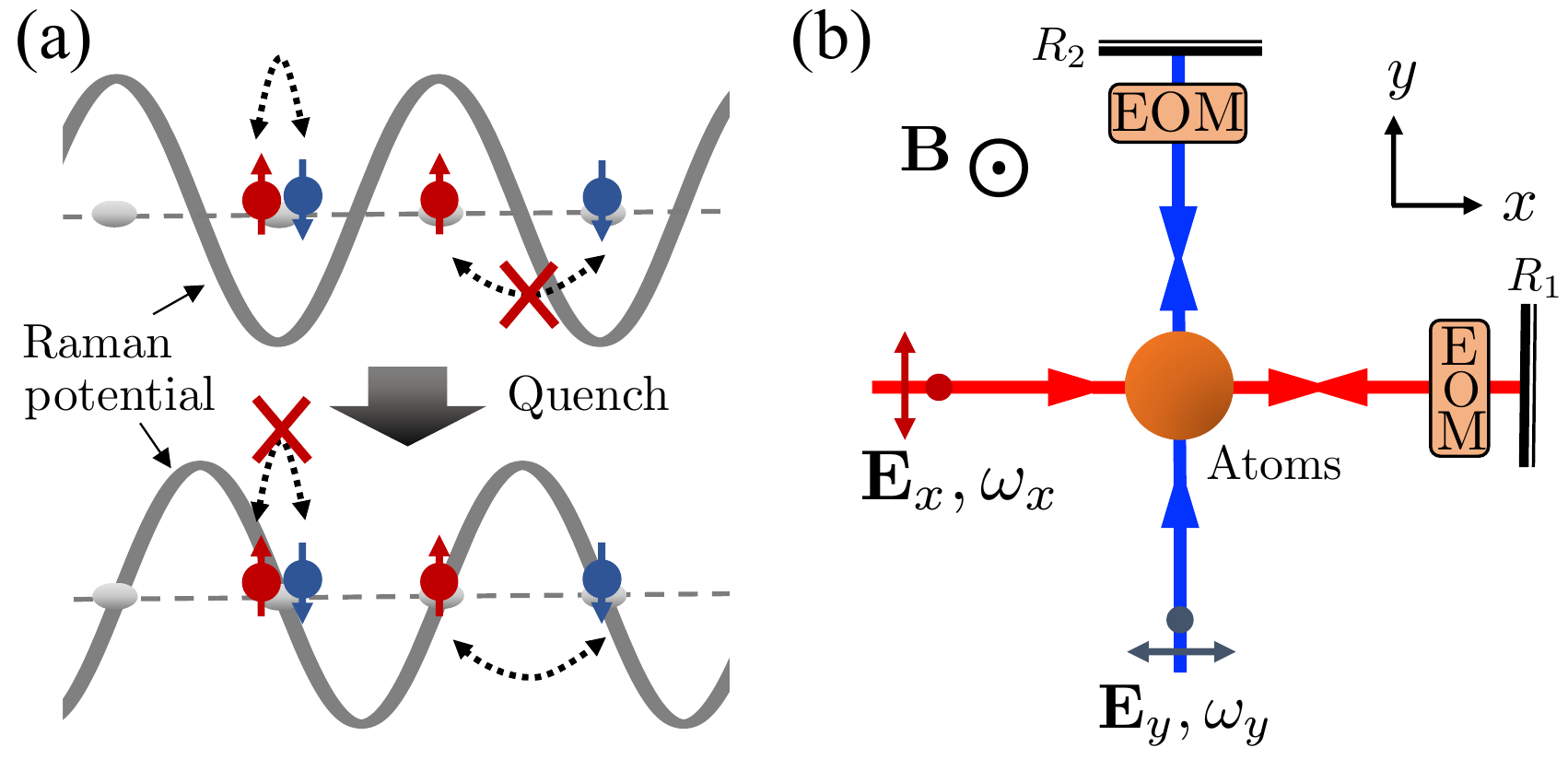}
\caption{
The realization of dynamical detection with optical Raman lattice.
(a) The process of quenching $h_x$ or $h_y$ can be realized by changing the relative symmetry of the Raman potential with respect to the lattice sites.
Before the quench, on-site spin-flipping is permitted, generating the constant magnetization $m_x\sigma_x$ or $m_y\sigma_y$;
after the quench, spin-flipped hopping is produced, realizing the QAH model.
(b) Experimental setup. A pair of laser beams ${\bf E}_x$ and ${\bf E}_y$, reflected by two mirrors $R_{1,2}$,
 produce square lattice and two independent Raman potentials for atoms.
Two EOMs are used to manipulate the relative symmetry of Raman potentials and realize the quench process.
}\label{fig4}
\end{figure}

{\em Experimental realization.---}
Now we propose an experimental setup (Fig.~\ref{fig4}) to detect topological charges in a 2D QAH model~\cite{XJLiu2014}, which has been realized in Refs.~\cite{Sun2018,Wu2016,Sun2017}.
This setup exploits the so-called optical Raman lattice scheme~\cite{Zhang_book}: two beams ${\bf E}_{x,y}$ to generate both 2D lattice $V_{\rm latt}(x,y)$ and Raman potentials ${\cal M}_{x,y}(x,y)$ simultaneously.
The quench process is realized by manipulating the relative symmetries between Raman and lattice potentials [Fig.~\ref{fig4}(a)].


First, the 2D QAH model can be realized as described in Ref.~\cite{Wang2018}.
Two electro-optic modulators (EOMs) are set to induce an additional $\pi/2$-phase shift for the $\hat{z}$-component field by manipulating the voltage [Fig.~\ref{fig4}(b)].
This gives the light fields
${\bf E}_x=\hat{y}E_{xy}\cos k_0 x+\ui\hat{z}E_{xz}\sin k_0x$
and ${\bf E}_y=\hat{x}E_{yx}\cos k_0 y+\ui\hat{z}E_{yz}\sin k_0y$,
where $E_{\mu\nu}$ ($\mu,\nu=x,y,z$) is the amplitude of the field in the $\mu$ direction and with the $\nu$ polarization.
For alkali metal atoms, the optical lattice potential is spin independent $V_{\rm latt}(x,y)=V_{0x}\cos^2 k_0x+V_{0y}\cos^2 k_0y$ (see Ref.~\cite{Supp} for optical transitions of $^{40}$K atoms), with the amplitudes $V_{0x}\propto(E_{xy}^2-E_{xz}^2)/\Delta$ and $V_{0y}\propto(E_{yx}^2-E_{yz}^2)/\Delta$.
One Raman potential ${\cal M}_x(x,y)=M_{0x}\cos k_0y\sin k_0x$, with
$M_{0x}\propto E_{xz}E_{yx}/\Delta$, is generated by the $E_{xz}$ and $E_{yx}$ components, and the other ${\cal M}_y(x,y)=M_{0y} \cos k_0x\sin k_0y$
with $M_{0y}\propto E_{xy}E_{yz}/\Delta$ by the $E_{xy}$ and $E_{yz}$ components~\cite{Supp}.
The Raman ${\cal M}_{x(y)}$ and lattice potentials satisfy a relative antisymmetric configuration along the $x$ ($y$) direction,
which ensures that ${\cal M}_{x(y)}$ leads to spin-flipped hopping only along the $x$ ($y$) direction [see Fig.~\ref{fig4}(a)].
Hence, for the $s$-band, the Bloch Hamiltonian is~\cite{Supp}
${\cal H}({\bf q})={\bf h}({\bf q})\cdot{\bm\sigma}=[m_z-t_0(\cos q_xa+\cos q_ya)]\sigma_z+t_{\rm so}\sin q_xa\sigma_y+t_{\rm so}\sin q_ya\sigma_x$,
which is the QAH model.
Here $a$ is the lattice constant, $q_{x,y}$ are Bloch momenta, $m_z$ measures the two-photon detuning,
and the spin-conserved ($t_0$) and spin-flipped ($t_{\rm so}$) hopping coefficients are respectively determined by the lattice
and Raman potentials.

The quench for all the components  $h_{x,y,z}$ are proposed as follow.
Quenching $h_z$ can be easily achieved by varying the two-photon detuning via the bias magnetic field.
Quenching  $h_{x,y}$ requires the generation of constant magnetization $m_x\sigma_x$ or $m_y\sigma_y$, which
can be realized by turning off one of the EOMs.
Take quenching $h_y$ as an example. If turning off the EOM in the $x$ direction,
the beams form the fields ${\bf E}'_x=\hat{y}E_{xy}\cos k_0 x+\hat{z}E_{xz}\cos k_0x$
and ${\bf E}_y=\hat{x}E_{yx}\cos k_0 y+\ui\hat{z}E_{yz}\sin k_0y$, which produce the lattice potential $V_{\rm latt}(x,y)=V'_{0x}\cos^2 k_0x+V'_{0y}\cos^2 k_0y$,
with $V'_{0x}\propto(E_{xy}^2+E_{xz}^2)/\Delta$ and $V'_{0y}=V_{0y}$ given as before, and Raman potentials
${\cal M}_x(x,y)=0$ and ${\cal M}_y(x,y)=M_{0x}\cos k_0y\cos k_0x+M_{0y} \cos k_0x\sin k_0y$.
The $M_{0y}$-term of the Raman potential induces spin-flipped hopping along $y$ direction, corresponding to
$t_{\rm so}\sin q_ya\sigma_x$ in the model, while
the former term ($M_{0x}$) is symmetric with respect to each lattice site,
leads to on-site spin-flipping [see Fig.~\ref{fig4}(a)] and generates a constant
magnetization $m_y\sigma_y$~\cite{Supp}.
Similarly, turning off the EOM in the $y$ direction can generate the term $m_x\sigma_x$. Thus, the quenches regarding $h_{x,y}$ can be readily implemented by suddenly varying the relative phase of lights
via the EOMs. An initial magnetization $m_{y(x)}\sim20t_0$ is achieved in typical parameter conditions with
$|V'_{0x}|=|V'_{0y}|=6.3E_{\rm r}$, $M_{0x(0y)}=2.5E_{\rm r}$, $M_{0y(0x)}=0$ ($E_{\rm r}\equiv\hbar^2k_0^2/2m$)~\cite{Supp}.

{\em Conclusions.---} We have proposed a generic scheme to characterize the phases by
a dynamical topological invariant defined at the monopoles of the SO field. The scheme is applicable to topological phases of integer $\mathbb{Z}$ invariants and of any dimension, with 2D and 3D models being numerically examined. Besides, this method
exhibits explicit advantages for experimental investigation due to the simplification of measurement.
In the future, we expect to generalize our theory to broader classes of topological phases, including those
classified by $\mathbb{Z}_2$.

This work was supported by the National Key R\&D Program of China (2016YFA0301604), National Nature Science Foundation of China
(under grants No. 11674301 and No. 11761161003), and the Thousand-Young-Talent Program of China.

%

\pagebreak
\clearpage
\setcounter{equation}{0}
\setcounter{figure}{0}

\newtheorem*{definition}{Definition}

\renewcommand{\theparagraph}{\bf}
\renewcommand{\thefigure}{S\arabic{figure}}
\renewcommand{\theequation}{S\arabic{equation}}

\onecolumngrid
\flushbottom


\section{\large Supplementary Information}

\subsection{I. Characterize Topological Phases by Charges}

We consider the generic $d$-dimensional ($d$D) Hamiltonian
\begin{equation}
{\cal H}({\bf k})={\bf h}({\bf k})\cdot{\bm \gamma}=h_0({\bf k})\gamma_0+\sum_{i=1}^d h_i({\bf k})\gamma_i,
\end{equation}
where the matrices ${\bm\gamma}$  obey the Clifford algebra $\left\{ \gamma_{i},\gamma_{j}\right\} =2\delta_{ij}\mathbbm{1}$ ($i,j=0,1\dots,d$), and are of dimensionality $n_{d}=2^{d/2}$ (or $2^{(d+1)/2}$) if $d$ is even (or odd).
This model involve the minimal bands to open a topological gap for the $d$D topological phase.
Without loss of the generality, we choose $h_{0}(\mathbf{k})$ to characterize the band structure and the remaining $d$-components
$h_{i\neq0}(\mathbf{k})$ to depict the coupling between bands.
We denote these components by the spin-orbit (SO) field ${\bf h}_{\mathrm{so}}(\mathbf{k})\equiv(h_{1},\dots,h_{d})$.
We emphasize that the $\gamma$ matrices are constructed to satisfy the trace property ${\rm Tr}[\prod_{j=0}^d\gamma_j]=(-2\ui)^n$ for even $d=2n$
or ${\rm Tr}[\gamma\prod_{j=0}^d\gamma_j]=(-2\ui)^n$ for odd $d=2n-1$, with $\gamma=\ui^n\prod_{j=0}^d\gamma_j$ being the chiral matrix.
For example, in 2D we should have ${\rm Tr}[\gamma_0\gamma_1\gamma_2]=-2\ui$; if $\gamma_0=\sigma_z$, one should set $\gamma_1=\sigma_y$ and $\gamma_2=\sigma_x$
(it is just the case in our numerical calculations of the 2D model).

In Ref.~\cite{Zhang2018_S}, we have proved that the $d$D bulk topology can be characterized by a $(d-1)$D invariant defined on the band-inversion surfaces (BISs):
\begin{equation}\label{theorem_s}
{\cal W} =\sum_{j}\frac{\Gamma(d/2)}{2\pi^{d/2}}\frac{1}{\left(d-1\right)!}\int_{{\rm BIS}_{j}}\hat{{\bf h}}_{\mathrm{so}}\bigl(\ud\hat{{\bf h}}_{\mathrm{so}}\bigr)^{d-1},
\end{equation}
where $\Gamma(x)$ is the Gamma function, $\hat{{\bf h}}_{\mathrm{so}}(\mathbf{k})\equiv{\bf h}_{\mathrm{so}}(\mathbf{k})/|{\bf h}_{\mathrm{so}}(\mathbf{k})|$ denotes the unit SO field, $\hat{{\bf h}}_{\mathrm{so}}(\ud\hat{{\bf h}}_{\mathrm{so}})^{d-1}\equiv\epsilon^{i_{1}i_{2}\cdots i_{d}}\hat{h}_{\mathrm{so},i_{1}}\ud\hat{h}_{\mathrm{so},i_{2}}\wedge\cdots\wedge\ud\hat{h}_{\mathrm{so},i_{d}}$, with $\epsilon^{i_{1}i_{2}\cdots i_{d}}$ being the fully anti-symmetric tensor and $i_{1,2,\dots,d}\in\{1,2,\dots,d\}$, and `$\ud$' denotes the exterior derivative.
The above formula can also be written as
\begin{align}\label{topo_invar_s}
{\cal W}&=\sum_{i\in{\cal V}_{\rm BIS}}{\cal C}_i,
\end{align}
with
\begin{align}
{\cal C}_i&=\frac{\Gamma(d/2)}{2\pi^{d/2}}\int_{{\cal S}_{i}}\frac{1}{|{\bf h}_{\mathrm{so}}|^d}\sum_{i=1}^{d}\left(-1\right)^{i-1}{h}_{\mathrm{so},i}\ud{h}_{\mathrm{so},1}\wedge\cdots\wedge\widehat{\ud{h}_{\mathrm{so},i}}\wedge\cdots\wedge\ud{h}_{\mathrm{so},d}.
\end{align}
Here ${\cal S}_i$ denotes a $(d-1)$D surface enclosing the $i$th monopole ${\bf k}={\bm \varrho}_i$ with ${\bf h}_{\rm so}({\bm \varrho}_i)=0$, and the term with a big hat is omitted.
We further have
\begin{align}
{\cal C}_i&=\frac{\Gamma(d/2)}{2\pi^{d/2}}\int_{{\cal S}_{i}}\frac{J_{{\bf h}_{\rm so}}({\bm \varrho}_i)}{|{\bf h}_{\mathrm{so}}|^d}\sum_{i=1}^{d}\left(-1\right)^{i-1}{k}_{i}\ud{k}_{1}\wedge\cdots\wedge\widehat{\ud{k}_{i}}\wedge\cdots\wedge\ud{k}_{d}\nonumber\\
&={\rm sgn}[J_{{\bf h}_{\rm so}}({\bm \varrho}_i)]\frac{\Gamma(d/2)}{2\pi^{d/2}}\int_{{\cal S}_{i}}\frac{|J_{{\bf h}_{\rm so}}({\bm \varrho}_i)|}{|{\bf h}_{\mathrm{so}}|^d} \sum_{i=1}^{d}\left(-1\right)^{i-1}{k}_{i}\ud{k}_{1}\wedge\cdots\wedge\widehat{\ud{k}_{i}}\wedge\cdots\wedge\ud{k}_{d}\nonumber\\
&={\rm sgn}[J_{{\bf h}_{\rm so}}({\bm \varrho}_i)]\frac{\Gamma(d/2)}{2\pi^{d/2}}\int_{{\bf h}_{\rm so}({\cal S}_{i})}\left[\sum_{i=1}^{d}\left(-1\right)^{i-1}\frac{{h}_{\mathrm{so},i}}{|{\bf h}_{\mathrm{so}}|^d}\ud{h}_{\mathrm{so},1}\wedge\cdots\wedge\widehat{\ud{h}_{\mathrm{so},i}}\wedge\cdots\wedge\ud{h}_{\mathrm{so},d}\right].
\end{align}
Here $J_{{\bf h}_{\rm so}}({\bm \varrho}_i)\equiv \det\left[(\partial h_{\rm so,i}/\partial k_j)\vert_{{\bf k}={\bm \varrho}_i}\right]$ denotes the Jacobian determinant.
The integral in the last line gives the the area of the $(d-1)$D sphere ${\bf h}_{\rm so}({\cal S}_{i})$, i.e. ${\rm Vol}[S^{d-1}]=2\pi^{d/2}/\Gamma(d/2)$. Thus, we have
\begin{align}\label{ci_s}
{\cal C}_i&={\rm sgn}[J_{{\bf h}_{\rm so}}({\bm \varrho}_i)].
\end{align}

\begin{figure}
\includegraphics[width=0.4\textwidth]{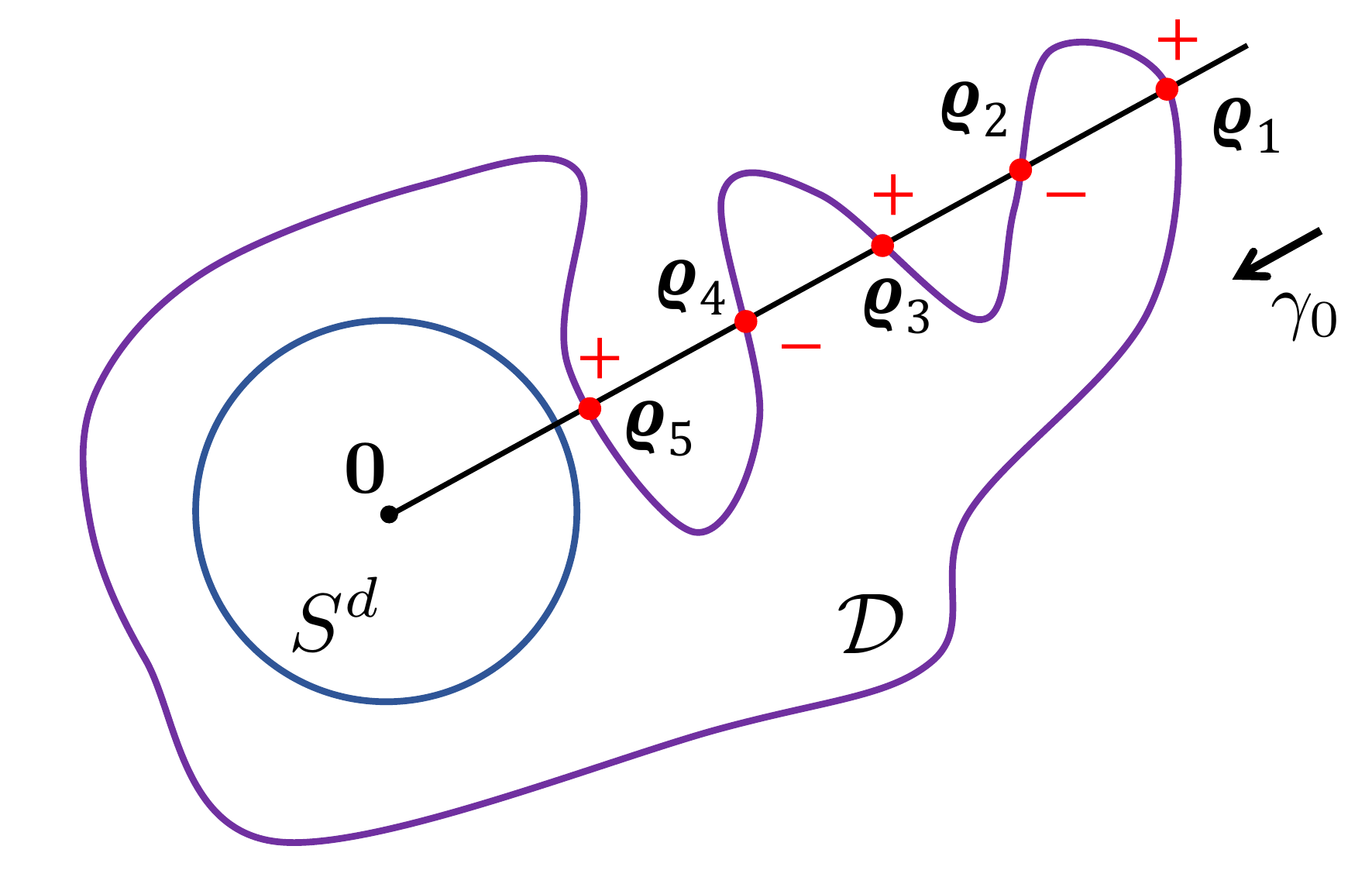}
\caption{Geometric interpretation of topological charges. The vector ${\bf h}(T^d)$ traces a closed parameter surface ${\cal D}$, passing through the negative $\gamma_0$ axis
at several interaction points ${\bm \varrho}_i$. These interactions are what we call `monopoles', with the charges being the the orientations `$\pm$'.
}\label{figS1}
\end{figure}

The repression (\ref{topo_invar_s}) using a summation of topological charges is directly related to the Brouwer degree~\cite{Felsager_book_S,Milnor_book_S,Sticlet2012_S}
of the mapping
$\hat{\bf h}({\bf k})\equiv{\bf h}({\bf k})/|{\bf h}({\bf k})|$.
We first introduce the definition of mapping degree here.
Let ${\cal M}$ and ${\cal N}$ be oriented $d$-dimensional manifolds without boundary and $f:{\cal M}\to{\cal N}$ be a smooth map. If ${\cal M}$ is compact and
${\cal N}$ is connected, then the degree of $f$ is defined as follows~\cite{Milnor_book_S}:
\begin{definition}
Let $x\in{\cal M}$ be a regular point of $f$, so that $df_x:T{\cal M}_x\to T{\cal N}_{f(x)}$ is a linear isomorphism between oriented tangent spaces. Define the sign of
$df_x$ to be $+1$ or $-1$ according as $df_x$ preserves or reverses orientation. For any regular value $y\in{\cal N}$ define the mapping degree
\begin{equation}
{\rm deg}(f;y)=\sum_{x\in f^{-1}(y)}{\rm sgn}(df_x).
\end{equation}
\end{definition}

Since ${\bf h}({\bf k})$ is nonzero for all {\bf k} in BZ, ${\bf h}: T^d\to\mathbb{R}^{d+1}\setminus\{{\bf 0}\}$ is a mapping from the Brillouin
zone torus $T^d$ to a $(d+1)$D closed surface ${\cal D}$. Thus $\hat{\bf h}$ is a composition
$\hat{\bf h}=\pi\circ{\bf h}$, where $\pi:\mathbb{R}^{d+1}\setminus\{{\bf 0}\}\to S^d$ is the central projection to the unit sphere.
According to the definition, for any regular point $z\in S^d$, the Brouwer degree is
\begin{equation}
{\rm deg}(\hat{\bf h}; z)=\sum_{{\bf k}\in \hat{\bf h}^{-1}(z)}{\rm sgn}[d\hat{\bf h}({\bf k})].
\end{equation}
The degree is the same for all regular points, so one can choose $z$ as a point in an axis of the unit sphere $S^d$, such as in the negative $\gamma_0$ axis. Denote
${\bf O}$ as the set of all the ${\bf k}$ points with $h_0({\bf k})<0$ and ${\bf h}_{\rm so}({\bf k})=0$.  We then have
\begin{equation}
{\rm deg}(\hat{\bf h})=\sum_{{\bm \varrho}_i\in {\bf O}}{\rm sgn}[J_{{\bf h}_{\rm so}}({\bm \varrho}_i)].
\end{equation}
We reobtain the expression of the topological invariant [Eqs.~(\ref{topo_invar_s}) and (\ref{ci_s})].
Therefore, as shown in Fig.~\ref{figS1}, the topological monopoles are the intersection points of the closed surface ${\cal D}$ and the negative $\gamma_0$ axis, and the charges
denote the orientation (or chirality) of the surface at each intersection.

\subsection{II.  Dynamical Detection of Topological Charges}
\subsubsection{A. General ideas}

We consider the spin dynamics of $\langle\gamma_0\rangle$ following different quenches. Each quench process is
realized by tuning $h_{i}(\mathbf{k})=m_i+\tilde{h}_i({\bf k})$ from a deep trivial regime with $m_i\gg{\rm max}[|\tilde{h}_i({\bf k})|]$ ($i>0$)  to
$0$ or tuning $m_0$ from
$m_0\gg{\rm max}[|\tilde{h}_0({\bf k})|]$ to a topological regime $m_0\lesssim{\rm max}[|\tilde{h}_0({\bf k})|]$ ($m_{i\neq0}=0$).
Here $\tilde{h}_i({\bf k})$ represents momentum-dependent field and $m_i$ demotes a constant magnetization.
The system with density matrix $\rho_i(0)$ is prepared in the ground state of the pre-quench Hamiltonian
in the deep trivial regime, where the spins are
fully polarized in the opposite direction of the $\gamma_{i}$ axis.
After a sudden quench, the quantum dynamics is governed by the unitary evolution operator $U(t)=\exp(-\ui\mathcal{H}t)$.
The time-dependent density matrix is then $\rho_i(t)=U(t)\rho_i(0)U^\dagger(t)$ and the spin texture is given by $\left\langle \gamma_{0}\right\rangle_i={\rm Tr}\left[\rho_i(t)\gamma_0\right]$.

The time-averaged spin polarization after quenching $h_i$ is measured as
\begin{equation}
\overline{\left\langle \gamma_{0}\right\rangle}_i=\lim_{T\to\infty}\frac{1}{T}\int_{0}^{T}\ud t\,\mathrm{Tr}\left[\rho_i(0)e^{\ui\mathcal{H}t}\gamma_{0}e^{-\ui\mathcal{H}t}\right].
\end{equation}
With the identity $e^{\ui\mathcal{H}t}=\cos(Et)+\ui\sin(Et)\mathcal{H}/E$, where $E(\bold k)=\sqrt{\sum_{i=0}^{d}h_{i}^{2}}$, we have
\begin{equation}
\overline{\left\langle \gamma_{0}\right\rangle}_i=\frac{h_0{\rm Tr}\left[\rho_i(0){\cal H}\right]}{E^2}.
\end{equation}
Since the initial state $\rho_i(0)$ is chosen such that ${\rm Tr}\left[\rho_i(0){\cal H}\right]=-h_i$,
the time averaged spin texture in the BZ is
\begin{eqnarray}\label{gamma0i_s}
\overline{\left\langle \gamma_{0}(\bold k)\right\rangle}_i=-h_{0}(\bold k)h_{i}(\bold k)/E^{2}(\bold k).
\end{eqnarray}
It is worthwhile to note that if we set the quench from $m_i\ll0$, the spin texture should be $\overline{\left\langle \gamma_{i}\right\rangle }=+h_{0}h_{i}/E^{2}$.

From the result, one can see that no matter which axis is quenched,
the time-averaged spin polarization $\overline{\left\langle \gamma_{0}(\bold k)\right\rangle }_i$ always vanishes on BISs.
Hence, a BIS can be dynamically determined by the surface with vanishing time-averaged spin polarization independent of the quench axis:
$\mathrm{BIS}=\{\mathbf{k}\vert\forall i\in\{0,1,2,\cdots,d\},\overline{\left\langle{\gamma_0}(\mathbf{k})\right\rangle}_i=0\}$.
In particular, $\overline{\langle\gamma_0\rangle}_0=0$ occurs only on BISs, so one can find out these surfaces simply by quenching $h_0$.
Besides, the time-averaged spin texture $\overline{\left\langle \gamma_{0}(\bold k)\right\rangle }_{i\neq0}$ also vanishes on the surface
with $h_i({\bf k})=0$ (see Eq.~\ref{gamma0i_s}). Accordingly, the monopoles of the SO field can be found out by $\overline{\left\langle \gamma_{0}({\bm \varrho})\right\rangle }_{i}=0$ for all $i>0$ but $\overline{\left\langle \gamma_{0}({\bm \varrho})\right\rangle }_{0}\neq0$. To characterize the charge, we find that near the point ${\bf k}={\bm\varrho}$, $E({\bf k})\approx|h_0({\bm\varrho})|$ and the time-averaged spin texture directly reflects the SO field:
\begin{equation}
\overline{\left\langle \gamma_{0}({\bf k})\right\rangle}_i\big\vert_{{\bf k}\to{\bm\varrho}}\simeq-h_{i}(\bold k)/h_{0}({\bm\varrho}).
\end{equation}
Thus, we define a dynamical field ${\bm \Theta}({\bf k})$, whose components are
\begin{equation}\label{fi_s}
\Theta_i({\bf k})\equiv-\frac{{\rm sgn}[h_0({\bf k})]}{{\cal N}_{\bf k}}\overline{\left\langle \gamma_{0}({\bf k})\right\rangle}_i,
\end{equation}
with ${\cal N}_{\bf k}$ being the normalization factor. It can be shown that near a topological monopole ${\bm\varrho}$, we have
\begin{equation}
\Theta_i({\bf k})\big\vert_{{\bf k}\to{\bm\varrho}}={h}_{{\rm so},i}({\bf k}).
\end{equation}
With this result, we reach immediately that the topological charge can be dynamically determined by
\begin{equation}
{\cal C}_i={\rm sgn}[J_{{\bm \Theta}}({\bm\varrho}_i)].
\end{equation}

\begin{figure}
\includegraphics[width=0.85\textwidth]{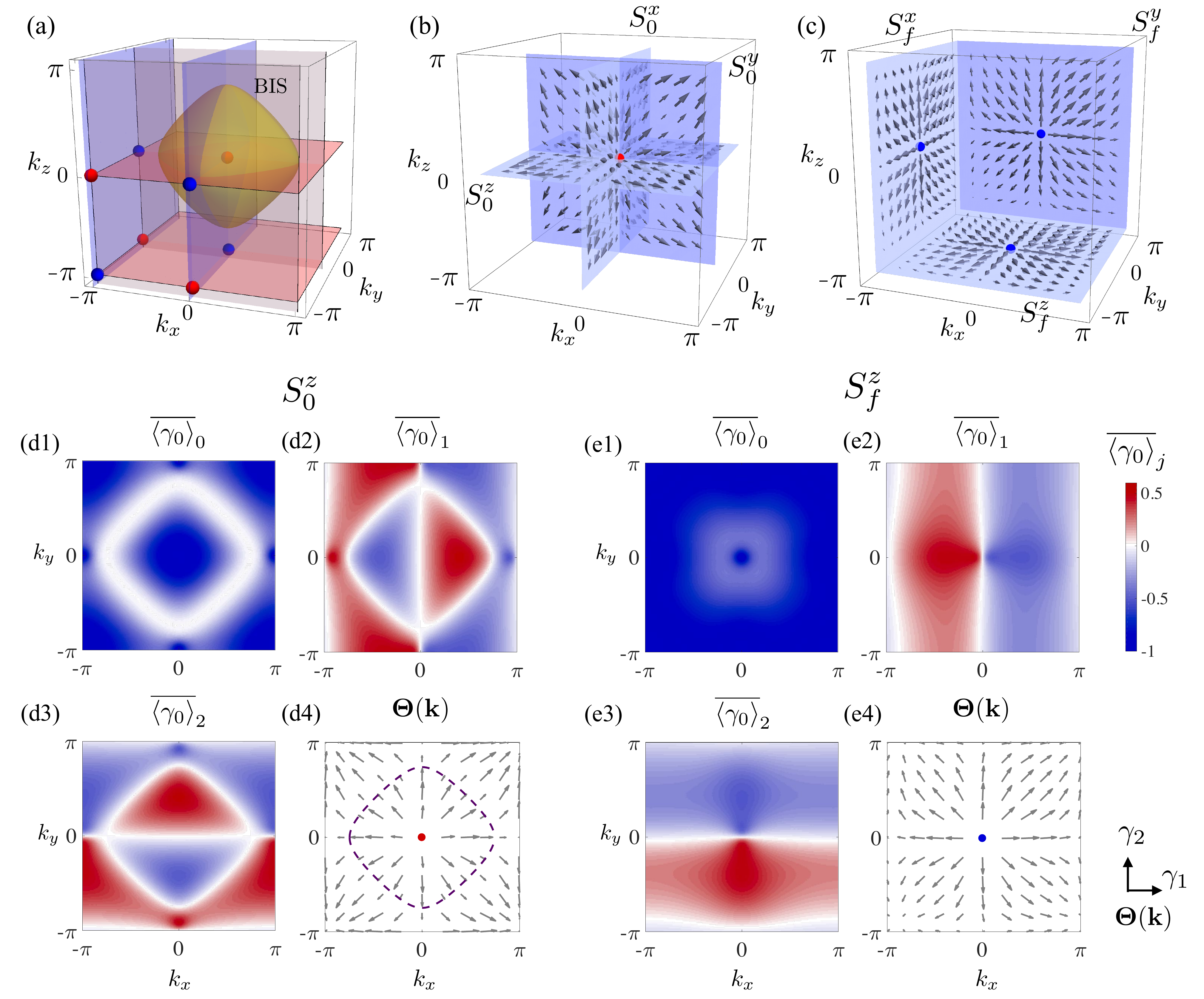}
\caption{Numerical results of the 3D model. (a) The observed time-averaged spin textures $\overline{\langle\gamma_0({\bf k})\rangle}_i=0$ determine the BIS and
the locations of topological monopoles. (b-c) Topological charges can be seen from the dynamical field on the $\overline{\langle\gamma_0\rangle}_i=0$ ($i=1,2,3$)
surfaces. These surfaces coincide with the planes $k_{x,y,z}=0$ (b) and $-\pi$ (c), denoted by $S_0^{x,y,z}$ and $S_f^{x,y,z}$, respectively. The dynamical field
takes the same pattern on $S_0^{x,y,z}$ or on $S_f^{x,y,z}$, and can be constructed by the spin textures, with an example shown in (d-e). (d-e) $S_0^{z}$ and
$S_f^{z}$ are the two planes with $\overline{\langle\gamma_0\rangle}_3=0$. While the spin texture $\overline{\langle\gamma_0({\bf k})\rangle}_0$ (d1,\,e1) can determine
the BIS, other two $\overline{\langle\gamma_0({\bf k})\rangle}_1$ (d2,\,e2) and  $\overline{\langle\gamma_0({\bf k})\rangle}_2$ (d3,\,e3) are used to construct the dynamical
field ${\bm \Theta}({\bf k})$ with the components $\Theta_{1,2}\propto\overline{\langle\gamma_0\rangle}_{1,2}$ (d4,\,e4). Here $t_{\rm so}=t_0$.
}\label{figS2}
\end{figure}

\subsubsection{B. Application to 3D topological phases}

Here we present the details of numerical calculations on the 3D model. The Hamiltonian reads $\mathcal{H}(\mathbf{k})={\bf h}(\mathbf{k})\cdot{\bm\gamma}$,
with $\gamma_{0}=\sigma_{z}\otimes\tau_{x}$, $\gamma_{1}=\sigma_{x}\otimes{\id}$,
$\gamma_{2}=\sigma_{y}\otimes{\id}$, $\gamma_{3}=\sigma_{z}\otimes\tau_{z}$, $h_{0}(\mathbf{k})=m_{0}-t_{0}\sum_{i}\cos k_{r_i}$ and
$h_{i}=m_i+t_{\mathrm{so}}\sin k_{r_i}$ [$i=1,2,3$ and $(r_1,r_2,r_3)=(x,y,z)$].
This model has a chiral symmetry $\gamma=\ui^2\gamma_0\gamma_1\gamma_2\gamma_3=-\sigma_z\otimes\tau_y$.
One can check that the constructed matrices satisfy the trace property
${\rm Tr}[\gamma\gamma_0\gamma_1\gamma_2\gamma_3]=(-2\ui)^2$.

In our calculations, the topological phase is set with the parameters $\{m_0,m_1,m_2,m_3\}=\{1.3t_0,0,0,0\}$, and one quench process is realized by tuning $m_i$ from $25t_0$ (others are all zero) to the set value ($m_0$ should be tuned to $1.3t_0$).
After the quench, the spin polarization will oscillate with time at each ${\bf k}$. The observed time-averaged spin textures can determine both the BIS
and the topological monopoles with $\overline{\langle\gamma_0({\bf k})\rangle}_i=0$ [see Fig.~\ref{figS2}(a)]: The surface with
$\overline{\langle\gamma_0({\bf k})\rangle}_0=0$ corresponds to the BIS (the green surface) and the
interaction points of $\overline{\langle\gamma_0({\bf k})\rangle}_{i>0}=0$ mark the locations of monopoles (colored dots).
Furthermore, based on the spin textures $\overline{\langle\gamma_0({\bf k})\rangle}_{i>0}$, we construct the dynamical field ${\bm \Theta}({\bf k})$ via the formula (\ref{fi_s}).
The results are shown in Fig.~\ref{figS2}(b-c), from which one can easily read out the charge of each monopole. As an example, we illustrate how to construct
the dynamical field on the planes $S_0^z$ ($k_z=0$) and $S_f^z$ ($k_z=-\pi$) in Fig.~\ref{figS2}(d-e).

\subsection{III. Experimental Realization for $^{40}$K Atoms}
\begin{figure}
\includegraphics[width=0.48\textwidth]{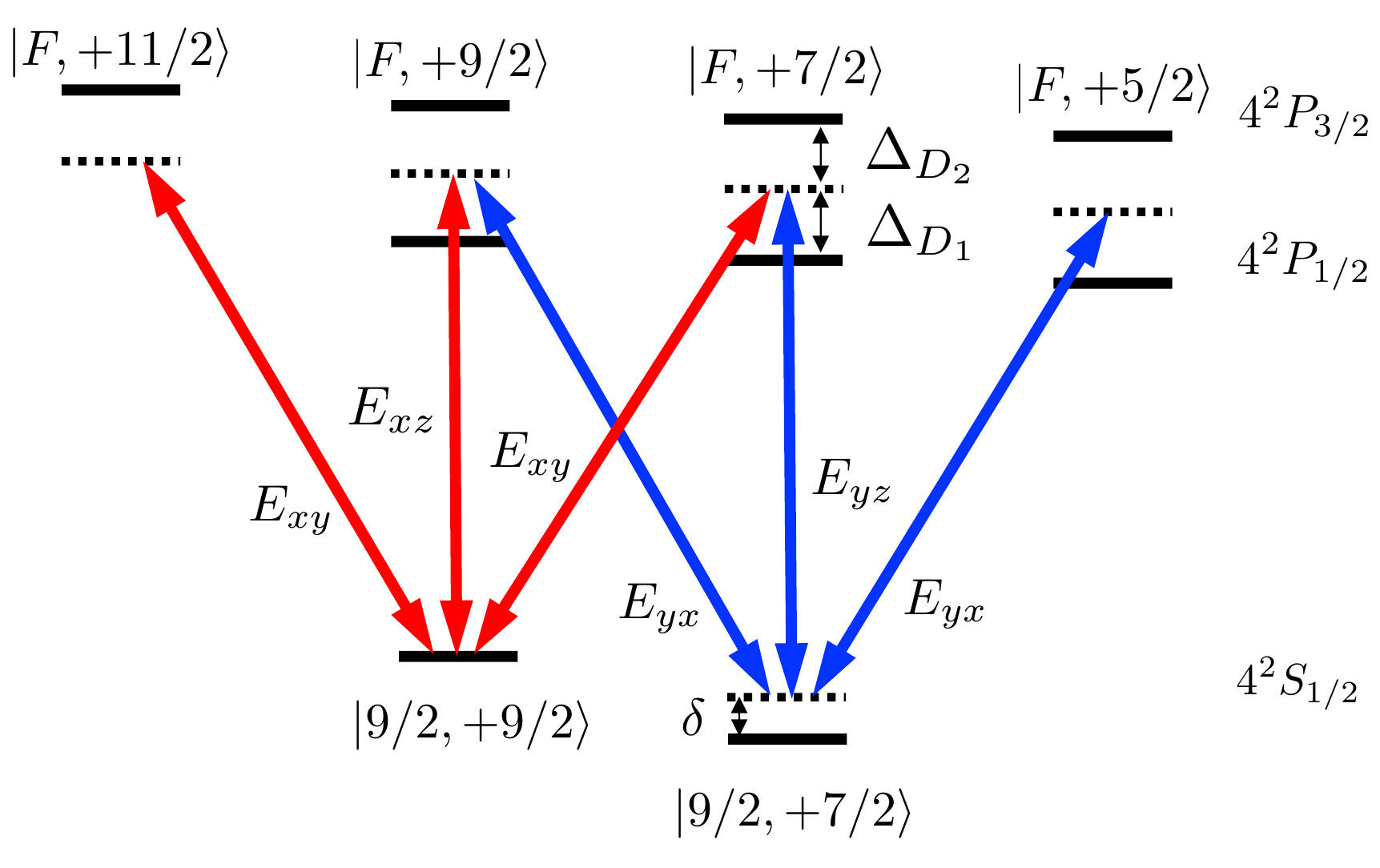}
\caption{Raman and lattice couplings for $^{40}$K atoms. Here the light beams are of wavelength between the $D_1$ and $D_2$ lines such that $\Delta_{D_1}>0$ and
$\Delta_{D_2}<0$.
}\label{figS3}
\end{figure}

Our experimental setup exploits the 2D optical Raman lattice, which has been realized in $^{87}$Rb bosons~\cite{Wu2016_S,Sun2017_S,Sun2018_S}.
Here, we take $^{40}$K atoms as an example while all our results are applicable to other alkali atoms. For $^{40}$K, the spin-$1/2$ system can be constructed
by $|g_{\uparrow}\rangle=|F=9/2, m_F=+9/2\rangle$ and $|g_{\downarrow}\rangle=|9/2, +7/2\rangle$. As shown in Fig.~\ref{figS3},
the lattice and Raman coupling potentials are contributed from both the $D_2$ ($4{^{2}S}_{1/2}\to4{^{2}P}_{3/2}$) and $D_1$ ($4^{2}S_{1/2}\to4^{2}P_{1/2}$) lines.
The total Hamiltonian reads ($\hbar=1$)
\begin{equation}\label{Ham_S}
H=\left[\frac{{\bf k}^2}{2m}+V_{\rm latt}(x,y)\right]\otimes{\bf 1}+{\cal M}_x(x,y)\sigma_x+{\cal M}_y(x,y)\sigma_y+m_z\sigma_z,
\end{equation}
where $V_{\rm latt}(x,y)$ denotes the square lattice potential, ${\cal M}_{x,y}(x,y)$ are the Raman coupling potentials, and $m_z=\delta/2$ measures the two-photon detuning of Raman coupling.

\subsubsection{A. 2D QAH model}
Details of the realization of a QAH model can be also found in Supplementary Material of Ref.~\cite{Wang2018_S}.
To generate 2D spin-orbit coupling, we use electro-optic modulators (EOMs) to introduce an additional $\pi/2$ phase for the $\hat{z}$-component field (see Fig.~4 of the main text).
Passing through a EOM twice leads to a $\pi$-phase shift.
Hence, the laser beams form the standing-wave fields for atoms (see Fig.~4 of the main text):
\begin{align}
{\bf E}_x&=\hat{y}E_{xy}e^{\ui(\alpha+\alpha_L/2)}\cos(k_0 x-\alpha_L/2)+\ui\hat{z}E_{xz}e^{\ui(\alpha+\alpha_L/2)}\sin(k_0 x-\alpha_L/2)\nonumber\\
{\bf E}_y&=\hat{x}E_{yx}e^{\ui(\beta+\beta_L/2)}\cos (k_0 y-\beta_L/2)+\ui\hat{z}E_{yz}e^{\ui(\beta+\beta_L/2)}\sin(k_0 y-\beta_L/2),
\end{align}
where $\alpha$ and $\beta$ denote the initial phases, and $\alpha_L$ ($\beta_L$) is the phase acquired by ${\bf E}_x$ (${\bf E}_y$) for an additional optical path
 back to the atom cloud (excluding the EOM effect).

 As shown in Fig.~\ref{figS3}, two independent Raman transitions are driven by the components ${\bf E}_{xz}$, ${\bf E}_{yx}$ and ${\bf E}_{xy}$, ${\bf E}_{yz}$, respectively,
 which leads to
 \begin{align}\label{M12_S}
M_1&=\sum_F\frac{\Omega^{(3/2)*}_{\uparrow F,xz}\cdot\Omega^{(3/2)}_{\downarrow F,yx}}{\Delta_{D_2}}+\sum_F\frac{\Omega^{(1/2)*}_{\uparrow F,xz}\cdot\Omega^{(1/2)}_{\downarrow F,yx}}{\Delta_{D_1}} \nonumber\\
 M_2&=\sum_F\frac{\Omega^{(3/2)*}_{\uparrow F,xy}\cdot\Omega^{(3/2)}_{\downarrow F,yz}}{\Delta_{D_2}}+\sum_F\frac{\Omega^{(1/2)*}_{\uparrow F,xy}\cdot\Omega^{(1/2)}_{\downarrow F,yz}}{\Delta_{D_1}},
 \end{align}
 where $\Delta_{D_1}>0$, $\Delta_{D_2}<0$, $\Omega^{(J)}_{\sigma F,\mu z}=\langle\sigma|ez|F,m_{F\sigma},J\rangle\hat{e}_z\cdot{\bf E}_{\mu z}$ and
 $\Omega^{(J)}_{\sigma F,\mu\nu}=\langle\sigma|ex|F,m_{F\sigma}+1,J\rangle\hat{e}_+\cdot{\bf E}_{\mu\nu}+\langle\sigma|ex|F,m_{F\sigma}-1,J\rangle\hat{e}_-\cdot{\bf E}_{\mu\nu}$ ($\mu,\nu=x,y$). Here $\hat{e}_+\cdot{\bf E}_{yx}=E_{yx}/\sqrt{2}$ and $\hat{e}_-\cdot{\bf E}_{xy}=\ui E_{xy}/\sqrt{2}$ represent the right- and left-handed components, respectively. From the dipole matrix elements of $^{40}$K~\cite{Tiecke_S}, we obtain
 \begin{align}
{\cal M}_x&=M_{0x}\sin(k_0 x-\alpha_L/2)\cos (k_0 y-\beta_L/2)\nonumber\\
{\cal M}_y&=M_{0y}\cos(k_0 x-\alpha_L/2)\sin(k_0 y-\beta_L/2),
 \end{align}
 where
\begin{equation}\label{M0xy_S}
M_{0x/0y}=\frac{t_{D_1}^2}{9}\left(\frac{1}{|\Delta_{D_1}|}+\frac{1}{|\Delta_{D_2}|}\right)E_{xz/xy}E_{yx/yz},
\end{equation}
with the transition matrix elements $t_{D_1}\equiv\langle J=1/2||e{\bf r}||J'=1/2\rangle$, $t_{D_2}\equiv\langle J=1/2||e{\bf r}||J'=3/2\rangle$ and $t_{D_2}\approx\sqrt{2}t_{D_1}$.
The optical lattice is given by
\begin{align}
V_{\sigma=\uparrow,\downarrow}&=\sum_F\frac{1}{\Delta_{D_2}}\left(\left|\Omega^{(3/2)}_{\sigma F,xz}\right|^2+\left|\Omega^{(3/2)}_{\sigma F,xy}\right|^2
+\left|\Omega^{(3/2)}_{\sigma F,yz}\right|^2+\left|\Omega^{(3/2)}_{\sigma F,yx}\right|^2\right) \nonumber\\
&+\sum_F\frac{1}{\Delta_{D_1}}\left(\left|\Omega^{(1/2)}_{\sigma F,xz}\right|^2+\left|\Omega^{(1/2)}_{\sigma F,xy}\right|^2
+\left|\Omega^{(1/2)}_{\sigma F,yz}\right|^2+\left|\Omega^{(1/2)}_{\sigma F,yx}\right|^2\right)\nonumber\\
&=V_{0x}\cos^2(k_0 x-\alpha_L/2)+V_{0y}\cos^2(k_0 y-\beta_L/2),
\end{align}
where
\begin{equation}
V_{0x/0y}=\frac{t_{D_1}^2}{3}\left(\frac{1}{|\Delta_{D_1}|}-\frac{2}{|\Delta_{D_2}|}\right)(E_{xy/yx}^2-E_{xz/yz}^2).
\end{equation}

In the tight-binding limit and only considering $s$-bands, the Hamiltonian (\ref{Ham_S}) takes the form in the momentum space~\cite{Wu2016_S}
$H=\sum_{\bf q} {\Psi}^\dagger_{\bf q}{\cal H}_{\rm f}{\Psi}_{\bf q}$,
where ${\Psi}_{\bf q}=({c}_{{\bf q}\uparrow},{c}_{{\bf q}\downarrow})$ and
\begin{equation}\label{Hq}
{\cal H}_{\rm f}=[m_z-t_0(\cos q_xa+\cos q_y a)]{\sigma}_z+t_{\rm so}\sin q_xa{\sigma}_y+t_{\rm so}\sin q_ya{\sigma}_x.
\end{equation}
Here  $a$ is the lattice constant, ${\bf q}$ is the Bloch momentum,
and  $t_0$ and $t_{\rm so}$ are, respectively, the spin-conserved and spin-flipped hopping coefficients
\begin{align}
t_0&=-2\int d{\bf r}\phi_{s}(x,y)\left[\frac{{\bf k}^2}{2m}+V_{\rm latt}({\bf r})\right]\phi_{s}(x-a,y),\nonumber\\
t_{\rm so}&=2M_{0}\int d{\bf r}\phi_{s}(x,y)\cos(k_0y)\sin(k_0x)\phi_{s}(x-a,y),
\end{align}
where $\phi_{s}(x,y)$ denotes the Wannier function and  $M_{0x}=M_{0y}=M_{0}$ is assumed.

\subsubsection{B. Hamiltonian for quenching $h_{x,y}$}

In our dynamical detection scheme, quenching $h_z$ can be easily achieved by varying the two-photon detuning $\delta$.
Quenching the other two components require producing the constant magnetization $m_x\sigma_x$ or $m_y\sigma_y$, which means that
on-site spin-flipping needs to be generated.  It can be verified that the transverse magnetization $m_x\sigma_x$ (or $m_y\sigma_y$) can be induced by
turning off the EOM in the $y$ (or $x$) direction [see Fig.~4(b) in the main text]. Here we take $m_y\sigma_y$ as an example.

When turning off the EOM in the $x$ direction, we have the light fields
\begin{align}
{\bf E}'_x&=\hat{y}E_{xy}e^{\ui(\alpha+\alpha_L/2)}\cos(k_0 x-\alpha_L/2)+\hat{z}E_{xz}e^{\ui(\alpha+\alpha_L/2)}\cos(k_0 x-\alpha_L/2),\nonumber\\
{\bf E}_y&=\hat{x}E_{yx}e^{\ui(\beta+\beta_L/2)}\cos (k_0 y-\beta_L/2)+\ui\hat{z}E_{yz}e^{\ui(\beta+\beta_L/2)}\sin(k_0 y-\beta_L/2),
\end{align}
where ${\bf E}_y$ is unchanged and the phases $\alpha$, $\beta$, $\alpha_L$ and $\beta_L$ are defined as before.
The lattice and Raman couplings are the same as in Fig.~\ref{figS3}. The Raman transitions $M_{1,2}$ [Eq.~(\ref{M12_S})] both contribute to
the $\sigma_y$ coupling.
Thus the Raman potentials read
\begin{align}
{\cal M}_x=0,\quad{\cal M}_y=M_{0x}\cos(k_0 x-\alpha_L/2)\cos(k_0 y-\beta_L/2)+M_{0y}\cos(k_0 x-\alpha_L/2)\sin(k_0 y-\beta_L/2),
\end{align}
with $M_{0x/0y}$ given by Eq.~(\ref{M0xy_S}). The lattice potential is
\begin{align}
V_{\sigma=\uparrow,\downarrow}=V'_{0x}\cos^2(k_0 x-\alpha_L/2)+V'_{0y}\cos^2(k_0 y-\beta_L/2),
\end{align}
with
\begin{equation}
V'_{0x}=\frac{t_{D_1}^2}{3}\left(\frac{1}{|\Delta_{D_1}|}-\frac{2}{|\Delta_{D_2}|}\right)(E_{xy}^2+E_{xz}^2)m,
\end{equation}
and $V'_{0y}=V_{0y}$.
One can see that in the Raman potential ${\cal M}_{y}$, the $M_{0x}$ term is symmetric with respect to each lattice site along $x$ or $y$ direction,
thus leading to on-site spin-flipping if only $s$-bands are considered. The other term ($M_{0y}$) keeps spin-flipped hopping in the $y$ direction.

Since
\begin{align}
\int d{\bf r}\phi_{s\uparrow}^{\vec{i}}({\bf r})\cos(k_0x)\sin(k_0y)\phi_{s\downarrow}^{\vec{j}}({\bf r})&=(-1)^{i_x+i_y}\int d{\bf r}\phi_{s\uparrow}({\bf r})\cos(k_0x)\sin(k_0y)\phi_{s\downarrow}({\bf r}-{\bf r}_{\vec j}+{\bf r}_{\vec i}).,\nonumber\\
\int d{\bf r}\phi_{s\uparrow}^{\vec{i}}({\bf r})\cos(k_0x)\cos(k_0y)\phi_{s\downarrow}^{\vec{i}}({\bf r})&=(-1)^{i_x+i_y}\int d{\bf r}\phi_{s\uparrow}({\bf r})\cos(k_0x)\cos(k_0y)\phi_{s\downarrow}({\bf r}),
\end{align}
the tight-binding Hamiltonian should take the form
\begin{eqnarray}
H&=&-\frac{t_0^x}{2}\sum_{\langle i_xj_x\rangle,\sigma}c^\dagger_{i_x\sigma}c_{j_x\sigma}-\frac{t_0^y}{2}\sum_{\langle i_yj_y\rangle,\sigma}c^\dagger_{i_y\sigma}c_{j_y\sigma}-\left[\sum_{i_y}(-1)^{i_x+i_y}\ui\frac{t^x_{\rm so}}{2}(c^\dagger_{i_x\uparrow}c_{i_x+1\downarrow}-c^\dagger_{i_x\uparrow}c_{i_x-1\downarrow})+{\rm h. c.}\right] \nonumber\\
&-&\ui m_y\sum_{\vec{i}}(-1)^{i_x+i_y}(c^\dagger_{\vec{i}\uparrow}c_{\vec{i}\downarrow}-c^\dagger_{\vec{i}\downarrow}c_{\vec{i}\uparrow})+\sum_{\vec{i}}m_z(n_{\vec{i}\uparrow}-n_{\vec{i}\downarrow}),
\end{eqnarray}
with 
\begin{align}
t^x_0&=-2\int d{\bf r}\phi_{s}(x,y)\left[\frac{{\bf k}^2}{2m}+V_{\rm latt}({\bf r})\right]\phi_{s}(x-a,y), \\
t^y_0&=-2\int d{\bf r}\phi_{s}(x,y)\left[\frac{{\bf k}^2}{2m}+V_{\rm latt}({\bf r})\right]\phi_{s}(x,y-a),\\
t^x_{\rm so}&=2M_{0y}\int d{\bf r}\phi_{s}(x,y)\cos(k_0x)\sin(k_0y)\phi_{s}(x-a,y),\\
m_y&=M_{0x}\int d{\bf r}\phi_{s}({\bf r})\cos(k_0x)\cos(k_0y)\phi_{s}({\bf r}).
\end{align}
After the transformation $c_{\vec{j}\downarrow}\to e^{i\pi{\bf r}_{\vec j}/a}c_{\vec{j}\downarrow}$, the Hamiltonian becomes
\begin{eqnarray}
H&=&-\frac{t_0^x}{2}\sum_{\langle i_xj_x\rangle}\left(c^\dagger_{i_x\uparrow}c_{j_x\uparrow}-c^\dagger_{i_x\downarrow}c_{j_x\downarrow}\right)
-\frac{t_0^y}{2}\sum_{\langle i_yj_y\rangle}\left(c^\dagger_{i_y\uparrow}c_{j_y\uparrow}-c^\dagger_{i_y\downarrow}c_{j_y\downarrow}\right)
-\left[\sum_{i_x}\frac{t^x_{\rm so}}{2}\ui(c^\dagger_{i_x\uparrow}c_{i_x+1\downarrow}-c^\dagger_{i_x\uparrow}c_{i_x-1\downarrow})+{\rm h. c.}\right] \nonumber\\
&-&\ui m_y\sum_{\vec{i}}(c^\dagger_{\vec{i}\uparrow}c_{\vec{i}\downarrow}-c^\dagger_{\vec{i}\downarrow}c_{\vec{i}\uparrow})+\sum_{\vec{i}}m_z(n_{\vec{i}\uparrow}-n_{\vec{i}\downarrow}),
\end{eqnarray}
The Fourier transform 
\begin{equation}\label{Fourier_S}
c_{{\vec j}\sigma}=\frac{1}{\sqrt{N}}\sum_{\bf q} e^{i{\bf q}\cdot{\bf r}_{\vec j}}c_{{\bf q}\sigma},
\end{equation}
with $N$ being the number of lattice sites,
yields the Bloch Hamiltonian
\begin{equation}\label{Hq}
{\cal H}_{\rm i}=[m_z-(t^x_0\cos q_xa+t_0^y\cos q_y a)]\sigma_z+t^x_{\rm so}\sin q_xa\sigma_x+m_y\sigma_y.
\end{equation}

Similarly, when turning off the EOM in the $y$ direction, one can generate the constant magnetization $m_x\sigma_x$.

\subsubsection{C. Quench steps and parameter estimation}

In summary, one can implement the quenches as follows:
(1) The EOMs are both tuned to introduce a $\pi/2$ phase shift and the system is governed by the Hamiltonian ${\cal H}_{\rm f}$ all the time.
Quench $h_z$ by suddenly changing the two-photon detuning $m_z$.
(2) Turn on only the EOM in the $y$ direction, and set $E_{xz},E_{yx},E_{xy},E_{yz}$ at desired values;
we have ${\cal H}_{\rm i}\approx m_y\sigma_y$. Turn on the other one such that both EOMs take effect; the system is described by ${\cal H}_{\rm f}$.
The process of quenching $h_x$ is thus accomplished.
(3) Similar to (2), the only difference is tuning on the EOM in the $x$ direction before both EOMs take effect.
This is equivalent to quenching $h_y$.
After each step of the three, one can observe the time evolution of spin polarization and obtain the time-averaged spin texture $\overline{\langle\sigma_z({\bf k})\rangle}_\alpha$ ($\alpha=x,y,z$).

Here we take quenching $h_y$ as an example for parameter estimation. The laser beams can be set at the wavelength of $768$nm (between the $D_1$ and $D_2$ lines~\cite{Tiecke_S}). The detunings are then $\Delta_{D_1}=(2\pi)1.07$THz, and  $\Delta_{D_2}=-(2\pi)0.66$THz, respectively.
Before the quench, one can set $E_{xz}=E_{yx}=E_0$ and $E_{xy}=E_{yz}\simeq0$, and have the results $V'_{0x}=V_{0y}=-V_0$,
$M_{0x}\approx0.39V_{0}$, and $M_{0y}\simeq0$. Here $V_0\equiv t_{D_1}^2E_0^2(2/|\Delta_{D_2}|-1/|\Delta_{D_1}|)/3$.
When $V_0=6.3E_{\rm r}$ with $E_{\rm r}\equiv k_0^2/(2m)$, we have $m_y\approx 20t_0$, which is large enough for magnetization.
After the quench, one should ensure $E_{xy}>E_{xz}$ and $E_{yx}>E_{yz}$ such that the lattice is kept red-detuned.
With proper parameters, such as $V_{0x/0y}=4E_{\rm r}$ and $M_{0x/0y}=1E_{\rm r}$, the lifetime of $^{40}$K degenerate gas can be $\sim150$ms~\cite{Wang2018_S}, long enough for experimental observation.


\begin{thebibliography}{99}%
\bibitem{Hasan2010} M. Z. Hasan and C. L. Kane, {\it Topological insulators}. Rev. Mod. Phys. \textbf{82}, 3045 (2010).

\bibitem{Qi2011} X. L. Qi and S. C. Zhang, {\it Topological insulators and superconductors}. Rev. Mod. Phys. \textbf{83}, 1057 (2011).

\bibitem{Su1980} W. P. Su, J. R. Schrieffer, and A. J. Heeger, {\it Soliton excitations in polyacetylene}. Phys. Rev. B \textbf{22}, 2099 (1980).

\bibitem{Atala2013} M. Atala, M. Aidelsburger, J. T. Barreiro, D. Abanin, T. Kitagawa, E. Demler, and I. Bloch, {\it Direct measurement of the Zak phase in topological Bloch bands}.
Nat. Phys. {\bf 9}, 795 (2013).

\bibitem{Liu2013} X.-J. Liu, Z.-X. Liu, and M. Cheng,
{\it Manipulating Topological Edge Spins in a One-Dimensional Optical Lattice}. Phys. Rev. Lett. {\bf 110}, 076401 (2013).

\bibitem{Song2018} B. Song, L. Zhang, C. He, T. F. J. Poon, E. Hajiyev, S. Zhang, X.-J. Liu, and G.-B. Jo,
{\it Observation of symmetry-protected topological band with ultracold fermions}. Sci. Adv. {\bf 4},
eaao4748 (2018).

\bibitem{Aidelsburger2013} M. Aidelsburger, M. Atala, M. Lohse, J. T. Barreiro, B. Paredes, and I. Bloch,
{\it Realization of the Hofstadter Hamiltonian with Ultracold Atoms in Optical Lattices.}
Phys. Rev. Lett. {\bf 111}, 185301 (2013).

\bibitem{Miyake2013} H. Miyake, G. A. Siviloglou, C. J. Kennedy, W. C. Burton, and W. Ketterle,
{\it Realizing the Harper Hamiltonian with Laser-Assisted Tunneling in Optical Lattices.}
Phys. Rev. Lett. {\bf 111}, 185302 (2013).

\bibitem{Jotzu2014} G. Jotzu, M. Messer, R. Desbuquois, M. Lebrat, T. Uehlinger, D. Greif, and T. Esslinger,
{\it Experimental realization of the topological Haldane model with ultracold fermions}. Nature {\bf 515}, 237 (2014).

\bibitem{Aidelsburger2015} M. Aidelsburger, M. Lohse, C. Schweizer, M. Atala, J. T. Barreiro, S. Nascimb\`ene, N. R. Cooper, I. Bloch, and N. Goldman,
{\it Measuring the Chern number of Hofstadter bands with ultracold bosonic atoms}. Nat. Phys. {\bf 11}, 162-166 (2015).

\bibitem{Wu2016} Z. Wu, L. Zhang, W. Sun, X.-T. Xu, B.-Z. Wang, S.-C. Ji, Y. Deng, S. Chen, X.-J. Liu, and J.-W. Pan,
{\it Realization of two-dimensional spin-orbit coupling for Bose-Einstein condensates.} Science {\bf 354}, 82 (2016).

\bibitem{Sun2017} W. Sun  B.-Z. Wang, X.-T. Xu, C.-R. Yi, L. Zhang, Z. Wu, Y. Deng, X.-J. Liu, S. Chen, and J.-W. Pan,
{\it Long-lived 2D Spin-Orbit coupled Topological Bose Gas.} arXiv:1710.00717.


\bibitem{quench1}  L. D'Alessio and M. Rigol, Dynamical preparation of Floquet Chern insulators. Nat. Commun. {\bf 6}, 8336 (2015).

\bibitem{quench2} M. D. Caio, N. R. Cooper, and M. J. Bhaseen, Quantum Quenches in Chern Insulators. Phys. Rev. Lett. {\bf 115}, 236403 (2015).

\bibitem{quench3} Y. Hu, P. Zoller, and J. C. Budich, Dynamical Buildup of a Quantized Hall Response from Nontopological States. Phys. Rev. Lett. {\bf 117}, 126803 (2016).

\bibitem{quench4} F. N. \"{U}nal, E. J. Mueller, and M. \"{O}. Oktel, Nonequilibrium fractional Hall response after a topological quench. Phys. Rev. A {\bf 94}, 053604 (2016).

\bibitem{quench5} J. H. Wilson, J. C.W. Song, and G. Refael, Remnant Geometric Hall Response in a Quantum Quench. Phys. Rev. Lett. {\bf 117}, 235302 (2016).


\bibitem{Flaschner2016} N. Fl\"aschner, B. S. Rem, M. Tarnowski, D. Vogel, D.-S. L\"uhmann, K. Sengstock, and C. Weitenberg,
{\it Experimental reconstruction of the Berry curvature in a Floquet Bloch band.} Science {\bf 352},1091 (2016).

\bibitem{Flaschner2018} N. Fl\"aschner, D. Vogel, M. Tarnowski, B. S. Rem, D.-S. L\"uhmann, M. Heyl, J. C. Budich, L. Mathey, K. Sengstock, and C. Weitenberg,
{\it Observation of dynamical vortices after quenches in a system with topology.} Nat. Phys. {\bf 14}, 265 (2018).

\bibitem{Wang2017} C. Wang, P. Zhang, X. Chen, J. Yu, and H. Zhai,
{\it Scheme to Measure the Topological Number of a Chern Insulator from Quench Dynamics}. Phys. Rev. Lett. {\bf 118}, 185701 (2017).

\bibitem{Zhang2018} L. Zhang, L. Zhang, S. Niu, X.-J. Liu, {\it Dynamical classification of topological quantum phases.} arXiv:1802.10061.

\bibitem{Tarnowski2017} M. Tarnowski, F. N. \"Unal, N. Fl\"{a}chner, B. S. Rem, A. Eckardt, K. Sengstock, and C. Weitenberg,
{\it Characterizing topology by dynamics: Chern number from linking number}. arXiv:1709.01046.

\bibitem{Sun2018} W. Sun, C.-R. Yi, B.-Z. Wang, W.-W. Zhang, B. C. Sanders, X.-T. Xu, Z.-Y. Wang, J. Schmiedmayer, Y. Deng, X.-J. Liu, S. Chen, and J.-W. Pan,
{\it Uncover Topology by Quantum Quench Dynamics.} arXiv:1804.08226.

\bibitem{Yu2018} J. Yu, {\it Measuring Hopf links and Hopf invariants in a quenched topological Raman lattice}. arXiv:180410358.

\bibitem{WYi2018} X. Qiu, T.-S. Deng, G. -C. Guo, and W. Yi, {\it Dynamical topological invariants and reduced rate functions for dynamical quantum phase transitions in two dimensions}. arXiv:1804.09032.

\bibitem{Zhou2018} L. Zhou and J. Gong, {\it Non-Hermitian Floquet topological phases with arbitrarily many real-quasienergy edge states}. arXiv:1807.00988.

\bibitem{Qiu2018} X. Qiu, T.-S. Deng, Y. Hu, P. Xue, and W. Yi, {\it Fixed points and emergent topological phenomena in a parity-time-symmetric
quantum quench}. arXiv:1806.10268.

\bibitem{KWang2018} K. Wang, X. Qiu, L. Xiao, X. Zhan, Z. Bian, W. Yi, and P. Xue, {\it Simulating dynamic quantum phase transitions in photonic quantum walks}. arXiv:1806.10871.

\bibitem{Supp} See Supplementary Information for details.

\bibitem{Chiu2013} C.-K. Chiu, H. Yao, and S. Ryu, {\it Classification of topological insulators and superconductors in the presence of reflection symmetry}. Phys. Rev. B \textbf{88}, 075142 (2013).

\bibitem{Chiu2016} C.-K. Chiu, J. C. Y. Teo, A. P. Schnyder, and S. Ryu, {\it Classification of topological quantum matter with symmetries}. Rev. Mod. Phys. \textbf{88}, 035005 (2016).

\bibitem{Haldane1988} F. D. M. Haldane, {\it Model for a Quantum Hall Effect without Landau Levels: Condensed-Matter Realization of the ``Parity Anomaly''}. Phys. Rev. Lett. \textbf{61}, 2015 (1988).

\bibitem{XJLiu2014} X. -J. Liu, K. T. Law, and T. K. Ng, {\it Realization of 2D Spin-Orbit Interaction and Exotic Topological Orders in Cold Atoms}. Phys. Rev. Lett. {\bf 112}, 086401 (2014).

\bibitem{Schnyder2008} A. P. Schnyder, S. Ryu, A. Furusaki, and A. W. W. Ludwig, {\it Classification of topological insulators and superconductors in three spatial dimensions}. Phys. Rev. B \textbf{78}, 195125 (2008).

\bibitem{Zhang2001} S. C. Zhang and J. Hu, {\it A four-dimensional generalization of the quantum Hall effect}. Science, {\bf 294}, 823 (2001).

\bibitem{Felsager_book}   B. Felsager, {\it Geometry, Particles, and Fields} (Springer Verlag, 1998).

\bibitem{Milnor_book} P. Milnor, {\it Topology from a Differential Viewpoint} (University Press of Virginia, 1965).

\bibitem{Sticlet2012} D. Sticlet, F. Pi\'echon, J.-N. Fuchs, P. Kalugin, and P. Simon, {\it Geometrical engineering of a two-band Chern insulator in two dimensions with arbitrary topological index.} Phys. Rev. B {\bf 85}, 165456 (2012).

\bibitem{Zhang_book} L. Zhang and X.-J. Liu, {\it Spin-orbit coupling and topological phases for ultracold atoms.} arXiv:1806.05628.

\bibitem{Wang2018} B.-Z. Wang, Y.-H. Lu, W. Sun, S. Chen, Y. Deng, and X.-J. Liu,
{\it Dirac-, Rashba-, and Weyl-type spin-orbit couplings: Toward experimental realization in ultracold atoms.}
Phys. Rev. A {\bf 97}, 011605(R) (2018).

\end{thebibliography}

\begin{thebibliography}{99}%
 \bibitem[S1]{Zhang2018_S} L. Zhang, L. Zhang, S. Niu, X.-J. Liu, {\it Dynamical classification of topological quantum phases.} arXiv:1802.10061.

 \bibitem[S2]{Felsager_book_S}   B. Felsager, {\it Geometry, Particles, and Fields} (Springer Verlag, 1998).

\bibitem[S3]{Milnor_book_S} P. Milnor, {\it Topology from a Differential Viewpoint} (University Press of Virginia, 1965).

\bibitem[S4]{Sticlet2012_S} D. Sticlet, F. Pi\'echon, J.-N. Fuchs, P. Kalugin, and P. Simon, {\it Geometrical engineering of a two-band Chern insulator in two dimensions with arbitrary topological index.} Phys. Rev. B {\bf 85}, 165456 (2012).

\bibitem[S5]{Wu2016_S} Z. Wu, L. Zhang, W. Sun, X.-T. Xu, B.-Z. Wang, S.-C. Ji, Y. Deng, S. Chen, X.-J. Liu, and J.-W. Pan, {\it Realization of two-dimensional spin-orbit coupling for Bose-Einstein condensates.} Science {\bf 354}, 82 (2016).

\bibitem[S6]{Sun2017_S} W. Sun  B.-Z. Wang, X.-T. Xu, C.-R. Yi, L. Zhang, Z. Wu, Y. Deng, X.-J. Liu, S. Chen, and J.-W. Pan, {\it Long-lived 2D Spin-Orbit coupled Topological Bose Gas.} arXiv:1710.00717.

\bibitem[S7]{Sun2018_S} W. Sun, C.-R. Yi, B.-Z. Wang, W.-W. Zhang, B. C. Sanders, X.-T. Xu, Z.-Y. Wang, J. Schmiedmayer, Y. Deng, X.-J. Liu, S. Chen, and J.-W. Pan,
{\it Uncover Topology by Quantum Quench Dynamics.} arXiv:1804.08226.

\bibitem[S8]{Wang2018_S} B.-Z. Wang, Y.-H. Lu, W. Sun, S. Chen, Y. Deng, and X.-J. Liu,
{\it Dirac-, Rashba-, and Weyl-type spin-orbit couplings: Toward experimental realization in ultracold atoms.}
Phys. Rev. A {\bf 97}, 011605(R) (2018).

 \bibitem[S9]{Tiecke_S} T. G. Tiecke, {\it Properties of Potassium} (2011). URL http://www.tobiastiecke.nl/archive/PotassiumProperties.pdf.
 \end{thebibliography}
\end{document}